\documentclass[review]{elsarticle}

\usepackage{times}
\usepackage{soul}
\usepackage{url}

\usepackage{graphicx}
\usepackage{subfigure}
\usepackage{amsmath}
\usepackage{amsfonts}
\usepackage{algorithm}
\usepackage{algorithmic}
\usepackage{multirow}
\usepackage{booktabs}
\usepackage{lineno}
\usepackage{pifont}
\usepackage{color}

\usepackage{hyperref}

\graphicspath{ {./} }

\begin{document}

\begin{frontmatter}
\title{MASS: Multi-task Anthropomorphic Speech Synthesis Framework}

\author[mymainaddress,mysecondaryaddress]{Jinyin Chen\corref{mycorrespondingauthor}}
\cortext[mycorrespondingauthor]{Corresponding author}
\ead{chenjinyin@zjut.edu.cn}

\author[mysecondaryaddress]{Linhui Ye}
\ead{yelinhui1019@163.com}

\author[myforthddress]{Zhaoyan Ming}
\ead{mingzhaoyan@gmail.com}

\address[mymainaddress]{Institute of Cyberspace Security, Zhejiang University of Technology, Hangzhou, China}
\address[mysecondaryaddress]{College of Information Engineering, Zhejiang University of Technology, Hangzhou, China}
\address[mythirdaryaddress]{College of Computer Science and Technology, Zhejiang University, Hangzhou, China}
\address[myforthddress]{Innovative Computing Center, Zhejiang Univeristy, Hangzhou, China}

\begin{abstract}
Text-to-Speech (TTS) synthesis plays an important role in human-computer interaction.  Currently, most TTS technologies  focus on the naturalness of speech, namely, making the speeches sound like humans. However, the key tasks of the expression of emotion and the speaker identity are ignored, which limits the application scenarios of TTS synthesis technology. To make the synthesized  speech more realistic and expand the application scenarios, we propose a multi-task anthropomorphic speech synthesis framework (MASS), which can synthesize speeches from text with specified emotion and speaker identity. The MASS framework consists of a base TTS module and two novel voice conversion modules: the emotional voice conversion module and the speaker voice conversion module. We propose deep emotion voice conversion model (DEVC) and deep speaker voice conversion model (DSVC) based on convolution residual networks. It solves the problem of feature loss during voice conversion. The model trainings are independent of parallel datasets, and are capable of many-to-many voice conversion. In the emotional voice conversion, speaker voice conversion experiments, as well as the multi-task speech synthesis experiments, experimental results show  DEVC and DSVC convert speech effectively. The quantitative and qualitative evaluation results of multi-task speech synthesis experiments show MASS can effectively synthesis speech with specified text, emotion and speaker identity.
\end{abstract}

\end{frontmatter}

\section{Introduction\label{Introduction}}
Text-to-speech (TTS) synthesis technology can synthesize speech based on the input text and has been applied to many fields, such as virtual anchors, voice news broadcasts, and automated story reading. For example, iFLYTEK combines this technology with image processing techniques and machine translation technology to launch virtual anchors; this technology is also applied in various voice assistants, such as Apple's Siri, Microsoft's Cortana, etc., which allows mobile devices to communicate with people; in the field of post and telecommunications, the technology has been used to realize electronic customer service, and complete services such as call charge inquiry and call charge payment with customers.

Although the current TTS models can synthesize high quality speech~\cite{wang2017tacotron,peng2019parallel,ren2019fastspeech,shen2018natural,li2019neural,arik2017deep,miao2020flow,chung2019semi,saito2019vocoder,rebai2015text}, usually an input text is mapped to the corresponding speech without emotion and speaker identity. However, in certain circumstance, this kind of speech is not anthropomorphic enough to adapt to scenes, for example story automatic reading, news broadcast, etc. It is essential to reflect emotion or speaker's identity, i.e. children like their mum to read the story books in bedtime. Aiming at this problem, we not only complete the mapping of text to speech, but also make the synthetic speech with the specified emotion and speaker identity, which is defined as multi-task anthropomorphic speech synthesis. In addition, we define the speech synthesis of specified text and emotion or specified text and speaker as dual-task speech synthesis, while the synthesis of specified text, or specified emotion, or specified speaker is defined as single-task speech synthesis.

TTS systems and voice conversion are similar to some extent, because their purpose is to generate high-quality voice with appropriate linguistic content. Therefore, some work combines the TTS model with the voice conversion model to realize the joint training. For example, Huanget et al~\cite{huang2021pretraining} proposed pretraning techniques for sequence-to-sequence voice conversion. Specifically, they use the decoder in the pretrained TTS module to train a speech encoder and use the voice conversion dataset to fine-tuning the obtained pretrained speech encoder and decoder to obtain a voice conversion model. It solves the problem of training the voice conversion model when the dataset is insufficient. Besides, ~\cite{huang2019voice,luong2019bootstrapping, luong2020nautilus} combines the TTS model with the voice conversion model to solve the problem of difficult training of the voice conversion model when the amount of training data is insufficient. ~\cite{zhang2019joint, zhang2020transfer, park2020cotatron, jia2018transfer} makes the voice conversion module and the TTS module share the decoder to improve the voice conversion model's performance. The purpose of these works is mainly to solve the problem of voice conversion model training when the amount of data is insufficient and improve the voice conversion model's conversion effect by using TTS model. Unlike these works, the MASS framework proposed in this paper combines the TTS model, the emotional voice conversion model and the speaker voice conversion model. It can synthesize the speech with specified text, emotion and speaker identity. The contribution of these works in solving the difficulty of training voice conversion models with insufficient training data and improving voice conversion models' performance is clear. However, these methods still have some problems: 1. the training of some models still requires parallel data sets~\cite{zhang2019joint, huang2019voice}; 2. some methods can only achieve one-to-one or many-to-one voice conversion~\cite{zhang2019joint, zhang2020transfer, luong2019bootstrapping, luong2020nautilus}; 3. the joint training method has an impact on the performance of TTS~\cite{zhang2019joint}; 4. reference audio is needed in the synthesis stage~\cite{zhang2020transfer, jia2018transfer}. These problems bring difficulties to multi-task speech synthesis.

 As for dual-task speech synthesis methods, Gibiansky et al~\cite{gibiansky2017deep} proposed a multi-speaker TTS synthesis method by applying low dimensional speaker embedding vector into the model. There are several emotional speech synthesis methods. For example, emotional speech synthesis methods based on reference audio feature embedding~\cite{gururani2019prosody,henter2018deep}, variational auto-encoder (VAE) and normalizing flows~\cite{aggarwal2020using}, etc. Furthermore, Lee et al.~\cite{lee2017emotional} proposed emotional end-to-end neural speech synthesizer based on Tacotron. And Li et al.~\cite{li2018emphasis} proposed EMPHASIS that is an emotional phoneme-based acoustic model. However, these methods can only synthesize speech with specified text and speaker identity or specified text and emotion, which are dual-task speech synthesis. Therefore, these methods still can not adapt to the above multi-task  anthropomorphic speech synthesis scenarios.

Address to the problem, multi-task speech synthetic methods based on reference audio feature embedding are proposed~\cite{skerry2018towards,wang2018style,valle2020mellotron,whitehill2019multi,bian2019multi}, which could synthesize speech with specified text, emotion and speaker identity. However, almost all of these methods need reference audio to synthesize the target speech. More specifically, they transfer emotion and speaker identity from reference audio to target speech. These methods are inconvenient to use in some scenarios. For example, in the field of autonomous driving, it is often inconvenient to obtain reference audio. As a result, multi-task anthropomorphic speech synthesis methods without reference audio are designed, for instance, variational auto-encoder (VAE)~\cite{hsu2018hierarchical}. However, the training of the model requires the use of audiobooks corpus. In other words, the dataset required for the training of the model needs to contain text annotations, multiple emotional expressions and multiple speaker identities at same time, called a multi-task dataset. This dataset can be both expensive and time consuming to record.

In view of the above problems, we proposed a multi-task anthropomorphic speech synthesis framework named MASS. The main contributions can be summarized as:

\begin{itemize}
\item  combine the TTS module, emotional voice conversion module and speaker voice conversion module in series to propose a multi-task anthropomorphic speech synthesis framework (MASS). As far as we know, this is the first method to achieve multi-task speech synthesis by connecting multiple modules in series. It can synthesize anthropomorphic speech without any referenced audio. The MASS framework training only needs single-task datasets, that is, the dataset used does not need to contain text annotations, multiple emotions and speaker identities at the same time, but only one of them is required.
\item The series connection allows the training between the modules to be carried out separately, so that each module does not affect each other. When the MASS framework performs a single-task TTS, the voice conversion module will not reduce the TTS performance. And the series combination makes the MASS framework flexible. We can replace the corresponding module with a better performance module to improve the MASS speech synthesis effect. This brings convenience to the improvement of speech synthesis effect in the future.
\item To reduce the loss of emotional features and speaker features in speech synthesis process, we propose a deep emotional voice conversion model (DEVC) and a deep speaker voice conversion model (DSVC) based on the convolution residual network. They can convert the speech synthesized by TTS and ensure the naturalness of the speech. Besides, their training don't rely on parallel datasets and can realize many-to-many voice conversion.
\item We conduct both qualitative and quantitative experiments to evaluate the naturalness of the speech synthesized by MASS and its similarity with the target speech's emotion and speaker identity. For DEVC and DSVC, we compare two state-of-the-art voice conversion models on the speech synthesized by the TTS model to verify the voice conversion effect.

\end{itemize}

The rest of the paper is organized as follows. In Section~\ref{Related works}, we discuss related work. Section~\ref{Method} describes the framework of MASS and the structure of DEVC and DSVC. Section~\ref{Experiments and Analysis} presents experimental evalutions. Section~\ref{Conclusion} concludes this paper.

\section{Related Works\label{Related works}}

\subsection{Text-to-Speech Model}
TTS synthesis technology is an essential technology for human-computer interaction, aiming at synthesizing understandable and natural speech that is indistinguishable from human speech based on the input text. In the early TTS synthesis methods, concatenative speech synthesis methods~\cite{hunt1996unit,campbell1997prosody,iida2003speech,lee2002segmental} select pre-recorded speech units from a designated speech library according to the information analyzed based on input text, and after necessary adjustments, concatenate them together to achieve speech synthesis. However, the speech synthesized by this method has the problem of phonation discontinuity between consecutive speech units. Statistical parametric speech synthesis is another early speech synthesis methods~\cite{tokuday2015directly,qian2014training,ren2008statistical,ze2013statistical,khodabakhsh2017spoofing,parthasarathy1992automatic}. Statistical parameter model extracts acoustic features from speech, such as fundamental frequency (F0), spectrum features, etc., and then carries out acoustic modeling on the extracted speech features to complete speech synthesis. However, it is difficult for statistical parameter model to reproduce the details of the speech, the synthesized speech usually sounds dull and unnatural.

Due to the excellent performance of deep neural network in many intellgient computing tasks, TTS synthesis technology has evolved from early speech synthesis methods to speech synthesis using deep neural network. Arik et al.~\cite{arik2017deep} split the early parametric speech synthesis model into multiple sub-modules, and replaced each sub-module with a neural network to proposed DeepVoice. The Tacotron~\cite{wang2017tacotron} model consists of an encoder and a decoder based on the attention mechanism, which can directly synthesize linear-scale spectrograms from text. It simplifies the training steps of the TTS model and is the first end-to-end speech synthesis model. Peng et al.~\cite{peng2019parallel} proposed the first non-autoregressive sequence-to-sequence structure, ParaNet structure, by using the fully convolution network structure. The Fastspeech model based on transformer and one-dimensional convolution structure is proposed by Ren et al~\cite{ren2019fastspeech}, which greatly improves the speed of speech synthesis. Shen et al.~\cite{shen2018natural} proposed Tacotron2 by using cyclic sequence-to-sequence feature prediction network and vocoder. Li et al.~\cite{li2019neural} used the multi-head attention mechanism to replace the RNN structure and attention mechanism in the Tacotron2 structure to further improve the quality of the synthetic speech.

\subsection{Speaker voice conversion}
Voice conversion is a technology that converts non-semantic information in speech while retaining semantic information, such as changing speaker identity and emotion in the speech. This technology has a wide range of application value and has been studied in many fields, such as pronunciation assistance~\cite{nakamura2012speaking}, speech enhancement~\cite{inanoglu2009data,turk2010evaluation,toda2012statistical}, etc.

In the early speaker voice conversion methods, Gaussian Mixture Model (GMM) can achieve speaker voice conversion by fitting the joint distribution of input features and output features~\cite{toda2005spectral,stylianou1998continuous,helander2010voice,erro2015interpretable}. During the process of voice conversion, the output features are derived based on the input features and GMM, but the features of the speech converted by GMM are often too smooth, the sense hearing of converted speech is unnatural and poor. Shuang et al.~\cite{shuang2006frequency} proposed a frequency warping method, which extracts formants from parallel speech data, uses paired formants to fit a piecewise linear warping function, and uses the warping function to expand and contract the spectral envelope of the speech. However, its drawback is that its modification to the spectrum envelope is too limited, even the height of the formant cannot be modified, and the voice conversion effect is poor.

The development of deep learning improves the efficiency and effect of speaker voice conversion. Deep neural network based on feed-forward~\cite{desai2010spectral,mohammadi2014voice,saito2017voice}, recurrent neural network~\cite{sun2015voice,nakashika2014voice,tobing2019voice}, sequence-to-sequence neural network~\cite{sutskever2014sequence,cho2014learning,bahdanau2014neural} and generative adversarial network (GAN)~\cite{kaneko2017sequence} have all realized success in speaker voice conversion task. However, the training of these models depends on parallel datasets, which is usually time and energy consuming to obtain ~\cite{kameoka2018stargan}.

It is challenging to train models with non-parallel datasets and obtain quality speaker voice conversion effect. Several speaker voice conversion technologies using non-parallel datasets have been proposed. Xie et al.~\cite{xie2016kl} proposed a method to complete speaker voice conversion based on automatic speech recognition system. Chou et al.~\cite{chou2019one} proposed the One-shot-VC method. In the voice conversion stage, only one sentence of the target speaker is needed to map the original speaker identity as the target speaker identity. In addition to the above methods, there are some methods that don't rely on paralle datasets, too. For example, speaker voice conversion technology based on auto-encoder~\cite{hsu2016voice,blaauw2016modeling,saito2018non,polyak2019attention}, bidirectional LSTM model ~\cite{sun2015voice}, and GAN~\cite{kaneko2017parallel,kameoka2018stargan,kaneko2019stargan,fang2018high}.

\subsection{Emotion voice conversion}
Emotional voice conversion is another direction of voice conversion technology. Kawanami et al.~\cite{kawanami2003gmm} first proposed a GMM-based spectrum conversion technology, but due to the limitation of GMM's inability to fit deep features, the converted spectrum is not broad enough to express the emotion of the target speech. Jianhua et al.~\cite{tao2006prosody} studied the F0 conversion method based on GMM and Classification and Regression Tree to map neutral prosody to emotional prosody. Tsuzuki et al.~\cite{tsuzuki2004constructing} and Yamagishi et al.~\cite{yamagishi2003modeling} used Hidden Markov Model (HMM) voice conversion framework to complete emotional voice conversion. Inanoglu et al.~\cite{inanoglu2009data} combined HMM, GMM and F0 selection methods to realize the conversion of F0, pronunciation duration and short-term spectrum respectively, but this method requires a lot of parallel data. All these emotion voice conversion methods cannot fully fit the deep features of the emotion in the speech, the effect of emotional voice conversion is often not ideal, and the quality of the converted speech is usually poor.

The development of deep learning improves the effect of emotional voice conversion. Many methods use wavelet transform to decompose the F0, and use deep neural network to model the decomposed F0 to realize speech emotion voice conversion~\cite{ming2016deep,ming2015fundamental,luo2016emotional,luo2017emotional,luo2019emotional,robinson2019sequence}. An exemplar-based voice conversion framework is also successful in emotional voice conversion. The emotional voice conversion can be achieved by constructing a joint exemplary that includes spectral features, aperiodic components, energy spectrum and scale-free F0~\cite{ming2016exemplar}. The voice conversion framework can simultaneously convert frequency spectrum, energy spectrum and F0.

However, the model training of the above emotional voice conversion methods all require parallel datasets. Zhou et al.~\cite{zhou2020transforming} modeled Mel-cepstral coefficients (MCC) and F0 respectively, and used CycleGAN to effectively convert MCC and F0 to realize the emotional voice conversion. Although the training of CycleGAN does not rely on parallel datasets, CycleGAN can only complete one-to-one emotional mapping. Recently, Zhou et al.~\cite{zhou2020converting} used VAW-GAN~\cite{hsu2017voice} to complete the emotional voice conversion. Unlike the previous work, the encoder in VAW-GAN can extract the independent emotion representation in the speech, so that the model can complete the conversion of any one's emotion. In addition, StarGAN~\cite{choi2018stargan} has also been applied to emotional voice conversion~\cite{rizos2020stargan}, which can realize many-to-many voice conversion without parallel datasets.

\section{Methodology\label{Method}}
\begin{figure}[htbp]
  \centering
  \includegraphics[width=1\linewidth]{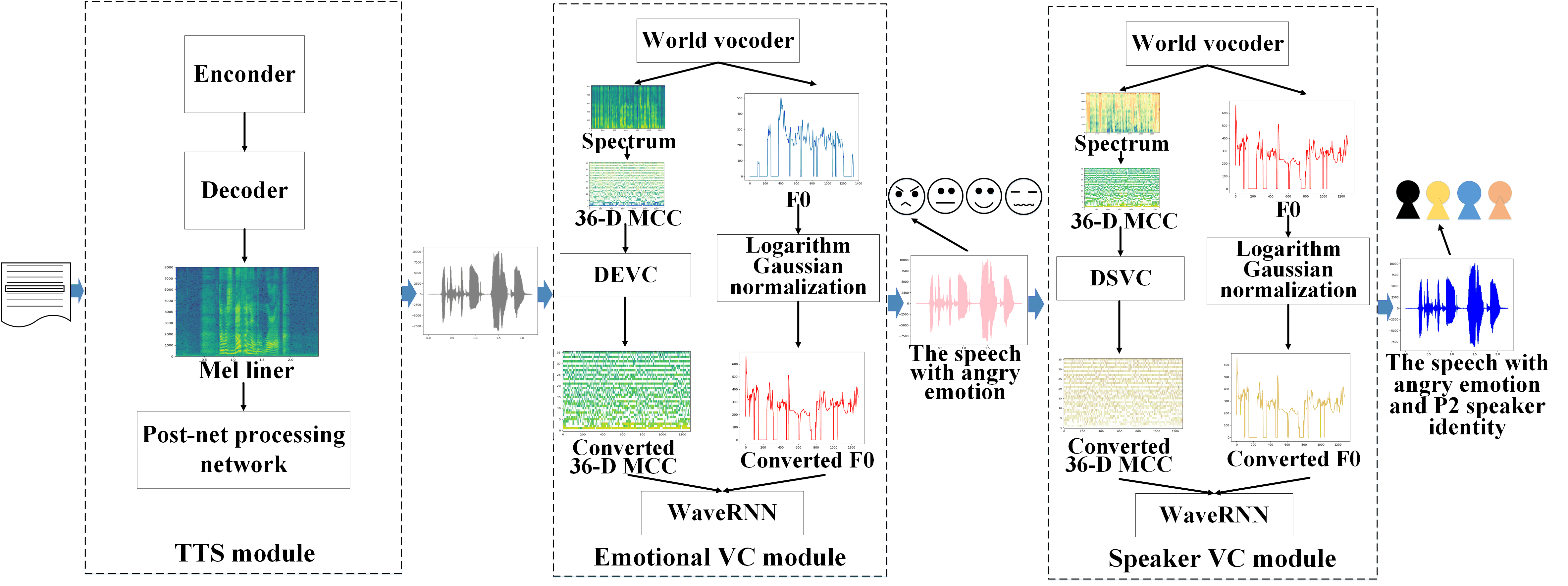}\\
  \caption{ Diagram of proposed MASS framework. The speech synthesized by TTS module is converted by emotional voice conversion module and speaker voice conversion module to achieve the multi-task anthropomorphic speech synthetic. The four expressions respectively represent emotions of angry, natural, amused and disgusted in synthesized speeches, and four color portraits represent four speaker identities in synthesized speeches.}
\label{fig:framework_overview}
\end{figure}

\subsection{System Overview\label{overview}}
The multi-task anthropomorphic speech synthesis framework proposed in this paper is shown in Figure~\ref{fig:framework_overview}. The whole MASS framework consists of three modules: TTS module, emotional voice conversion module and speaker voice conversion module. First, the TTS module synthesizes the speech with the specified text according to the input text. Then the emotional voice conversion module performs emotional voice conversion on the speech synthesized by the TTS module to synthesis target emotional speech. Finally, the speaker voice conversion module performs speaker voice conversion on the converted target emotional speech to obtain the target speaker's emotional speech. DEVC is trained on a dataset composed of multiple emotions of one speaker. Therefore, when DEVC performs emotional voice conversion, the output speech identity is the same as the emotional dataset's speech identity. Emotional voice conversion is performed on the speech through DEVC first. The speaker voice conversion is performed through DSVC to ensure that the synthesized speech has a specified emotion and identity. If the DEVC and DSVC are exchanged, it will not be possible to effectively synthesize the speech with the specified text, emotion and speaker identity. Therefore, we first convert the speech synthesized by TTS to the specified emotional speech through DEVC, and then convert it through DSVC to complete the multi-task speech synthesis process. As shown in Figure~\ref{fig:framework_overview}, during voice conversion, the World Vocoder~\cite{morise2016world} extracts the spectrum and F0 of the speech which need to be converted. Then convert the spectrum into 36-dimensional MCC, and perform emotional voice conversion and speaker voice conversion for MCC through DEVC and DSVC respectively. As for the F0, the Logarithm Gaussian normalization algorithm~\cite{liu2007high} is used for conversion. Finally, the target speech is reconstructed by the WaveRNN~\cite{kalchbrenner2018efficient}.

\subsection{Text-to-Speech module\label{module}}

The TTS module synthesizes the corresponding speech according to the input text. We choose the TTS-Transformer model~\cite{li2019neural} because its good performance and efficiency of training. This model uses a multi-head attention mechanism to replace the RNN and the attention mechanism between the encoder and the decoder in Tacotron2~\cite{shen2018natural}. On one hand, the multi-head attention mechanism can construct hidden states in the encoder and decoder in parallel to improve training efficiency. On the other hand, through the self-attention mechanism, any two words in the input can be connected, which solves the problem of the long-term dependence that RNN is difficult to model.

\subsection{WaveRNN\label{wavernn}}
We use WaveRNN as a vocoder to reconstruct speech. To ensure the synthesized speech's emotional and identity features, we concatenate MCC and F0 as feature vectors to train WaveRNN. In order to ensure that WaveRNN can synthesize high-quality speech without affecting the effect of multi-task speech synthesis, we have adopted different training strategies for WaveRNN in the emotional voice conversion module and speaker voice conversion module. First, in order to ensure the speech synthesis performance of the WaveRNN vocoder, we use the dataset that trains TTS module to train WaveRNN to ensure the quality of speech synthesis. For the WaveRNN in the emotional voice conversion module, we use the emotion dataset to fine-tuning it. The WaveRNN does not affect the emotional features in the process of synthesizing speech. For WaveRNN in speaker voice conversion, we further use the multi-speaker dataset and emotion dataset for fine-tuning so that WaveRNN can ensure that the emotional features and speaker features are not affected during the multi-task synthesis process.

\subsection{Emotional voice conversion module\label{Emotional VC module}}
Emotional voice conversion module changes the emotional patterns of the input speech, and outputs the emotional speech. It is composed of World vocoder and DEVC. World vocoder extracts MCC feature of the input speech, which will be converted by DEVC.

We are inspired by StarGAN-VC~\cite{kameoka2018stargan} to construct DEVC, in order to complete many-to-many emotional voice conversion. The training framework and the loss function are similar to StarGAN-VC. As shown in Figure~\ref{fig:DEVC}, the DEVC training framework consists of a generator, a discriminator and a classifier, which forms three adversarial networks.

\subsubsection{Training objectives}
\begin{figure}[htbp]
  \centering
  \includegraphics[width=0.85\linewidth]{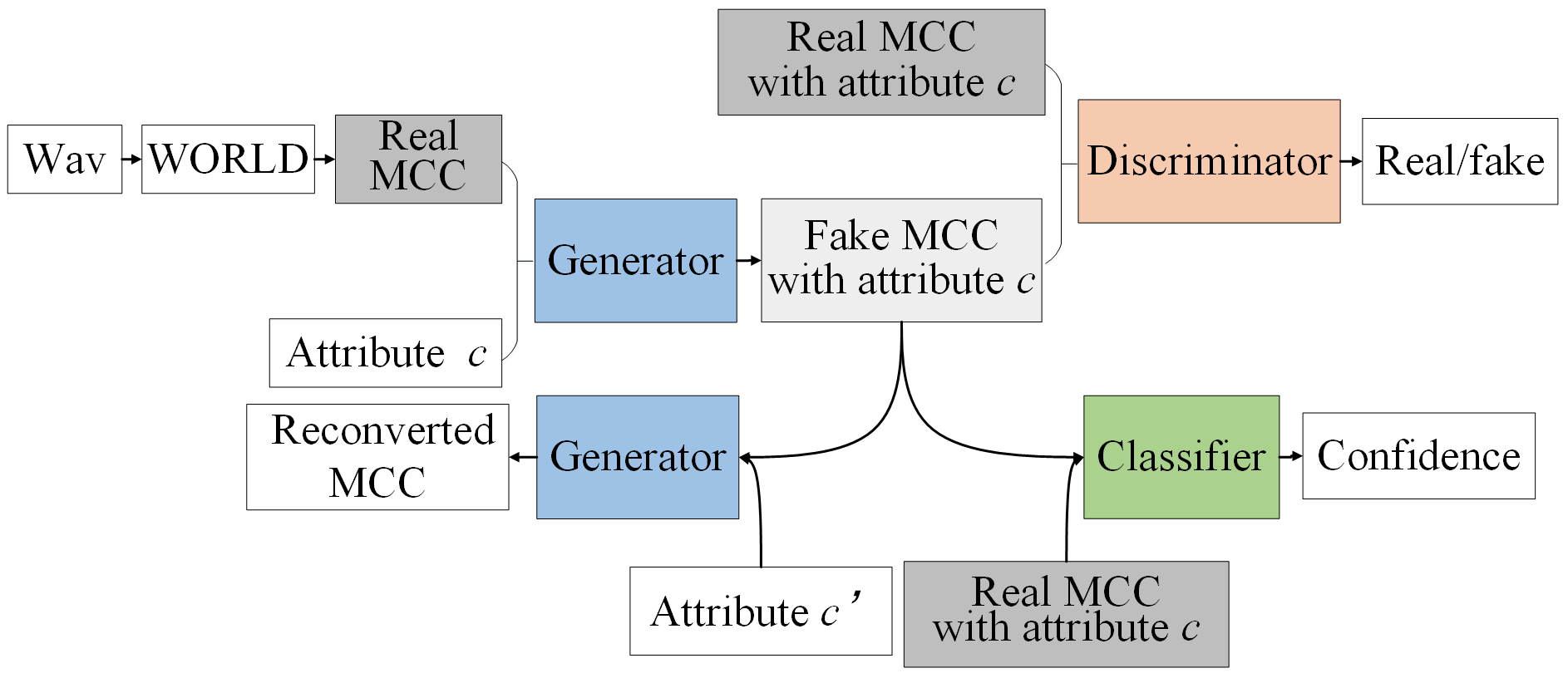}\\
  \caption{Diagram of DEVC training framework.}
\label{fig:DEVC}
\end{figure}

The training framework of DEVC is shown in the Figure~\ref{fig:DEVC}. We apply the convolution residual network to StarGAN-VC.

The first adversarial network is the adversary between the generator and the discriminator, which is the same as the general GAN structure.

The generator, the discriminator,  and the classifier in DEVC are constructed based on the convolution residual network. The optimization objectives of the generator and discriminator in the adversarial process are as follows:
\begin{small}
\begin{equation}
\begin{aligned}
\begin{gathered}
L _ { a d v } ^ { D } ( D ) = - E _ { c \sim p ( c ) , y \sim p ( y \mid c ) } [ \log D ( y , c ) ] - E _ { x \sim p ( x ) , c \sim p ( c ) } [ \log ( 1 - D ( G ( x , c ) , c ) ) ]\\
L _ { a d v } ^ { G } ( G ) = - E _ { x \sim p ( x ) , c \sim p ( c ) } [ \log D ( G ( x , c ) , c ) ]
\end{gathered}
\end{aligned}
\end{equation}
\end{small}
$ c \sim p ( c )$ denotes  the categories of emotions. $y\sim p(y\mid c )$ and $ x\sim p(x)$  represent the data with attribute $c$ in the training set and the data with any attribute in the training set, respectively. $L _ { a d v } ^ { D } ( D )$ is the training object of the discriminator, which enables the discriminator to learn whether the sequence generated by the generator is an acoustic feature sequence, i.e. MCC. Optimization objective $L _ { a d v } ^ { G } ( G )$ makes the generator have the ability to generate MCC.

The adversary between the generator and  the classifier forms the second adversarial network, which enables the generator to map the attributes of the original data to the specified target attributes, such as converting the speech of natural class to angry class. The optimal objectives of classifier and generator are as follows:
\begin{equation}
\begin{aligned}
\begin{gathered}
L _ { c l s } ^ { C } ( C ) = - E _ { c \sim p ( c ) , y \sim p ( y \mid c ) } [ \log C ( c \mid y ) ]  \\
L _ { c l s } ^ { G } ( G ) = - E _ { x \sim p ( x ) , c \sim p ( c ) } [ \log C ( c \mid G ( x , c ) ) ]
\end{gathered}
\end{aligned}
\end{equation}
$L _ { c l s } ^ { C } ( C )$ is the optimal quantitatives of the classifier, which makes the classifier categorize MCC according to the emotion.  $L _ { c l s } ^ { G } ( G )$  enables the generator to produce MCC with specified attributes.

The last adversarial network is the self-adversary of the generator. Only relying on the above two adversary networks cannot guarantee that the generator will not affect the semantic information in the extracted features. Therefore, the generator carries out self-adversary to retain the semantic information. The optimization objectives of generator self confrontation are as follows:
\begin{equation}
\begin{aligned}
\begin{gathered}
L _ { c y c } ( G ) = E _ { c ^ { \prime } \sim p ( c ) , x \sim p \left( x \mid c ^ { \prime } \right) , c \sim p ( c ) } \left[ \left\| G \left( G ( x , c ) , c ^ { \prime } \right) - x \right\| _ { \rho } \right] \\
L _ { i d } ( G ) = E _ { c ^ { \prime } \sim p ( c ) , x \sim p \left( x \mid c ^ { \prime } \right) } \left[ \left\| G \left( x , c ^ { \prime } \right) - x \right\| _ { \rho } \right]
\end{gathered}
\end{aligned}
\end{equation}
$x\sim p( x\mid c ^{\prime})$ represents the speech with $c ^ { \prime } $ attribute in the training dataset. $L _ { c y c } ( G )$ guarantees that the MCC generated by the generator has little difference from the original MCC when it is converted into the original attribute, so as to ensure that the generator will not affect the semantic information of speech. $L _ { i d } ( G )$ ensure that the generator will not affect the semantic information of the speech when converting between the same attributes.

Therefore, the optimization objectives of the generator, discriminator and classifier in DEVC are as follows:
\begin{equation}
\begin{aligned}
\begin{gathered}
L _ { G } = \lambda _ { a d v } L _ { a d v } ^ { G } + \lambda _ { c l s } L _ { c l s } ^ { G } + \lambda _ { c y c } L _ { c y c } ^ { G } + \lambda _ { i d } L _ { i d } ^ { G } \\
L _ { D } = L _ { a d v } ^ { D } \\
L _ { C } = L _ { c l s s } ^ { C }
\end{gathered}
\end{aligned}
\end{equation}
$\lambda _ { a d v }$, $\lambda _ { c l s }$, $\lambda _ { c y c }$ and $\lambda _ { i d }$ are hyperparameters greater than 0. In the training process, $\lambda _ { a d v }$ and $\lambda _ { c l s }$ should be relatively close, while $\lambda _ { c y c }$ and $\lambda _ { i d }$ should be smaller. This is to ensure that the generator can make the generated speech with the target emotion and ensure its naturalness.

\subsubsection{Generator network structure}
\begin{figure}[htbp]
  \centering
  \includegraphics[width=0.45\linewidth]{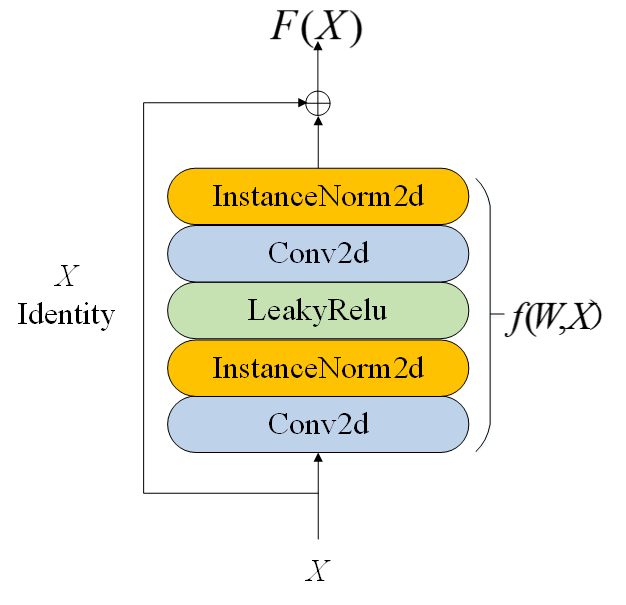}\\
  \caption{Diagram of CBlock}
\label{fig:CBlock}
\end{figure}

We apply the convolution residual network to the generator. The reasons are that, firstly it can achieve better voice conversion by extracting deep features, and will not cause the problem of gradient vanishing or explosion~\cite{he2016deep}. Secondly, it helps to solve the problem of feature loss in the voice conversion process, which will be elaborated in Section~\ref{Speaker VC module}. Figure~\ref{fig:CBlock} shows the convolution residual block (CBlock) we have constructed.

In CBlock, Conv2d means 2-dimensional convolution layer, which is used to extract features from the input. The extracted features are normalized by InstanceNorm2d. Instance Normalization was originally used for image style transfer~\cite{ulyanov2016instance}. Ulyanovd et al.~\cite{ulyanov2016instance} have proved that the mean and variance of each channel of the feature map output by the convolutional layer will affect the style of the final generated image, so normalizing the convolutional feature map on each channel will get better effect. The emotional voice conversion of speech is actually similar to the style conversion of images. The difference lies in the channel of the input generator. Therefore, we also use Instance Normalization to normalize the feature map.

\begin{figure}[htbp]
  \centering
  \includegraphics[width=1\linewidth]{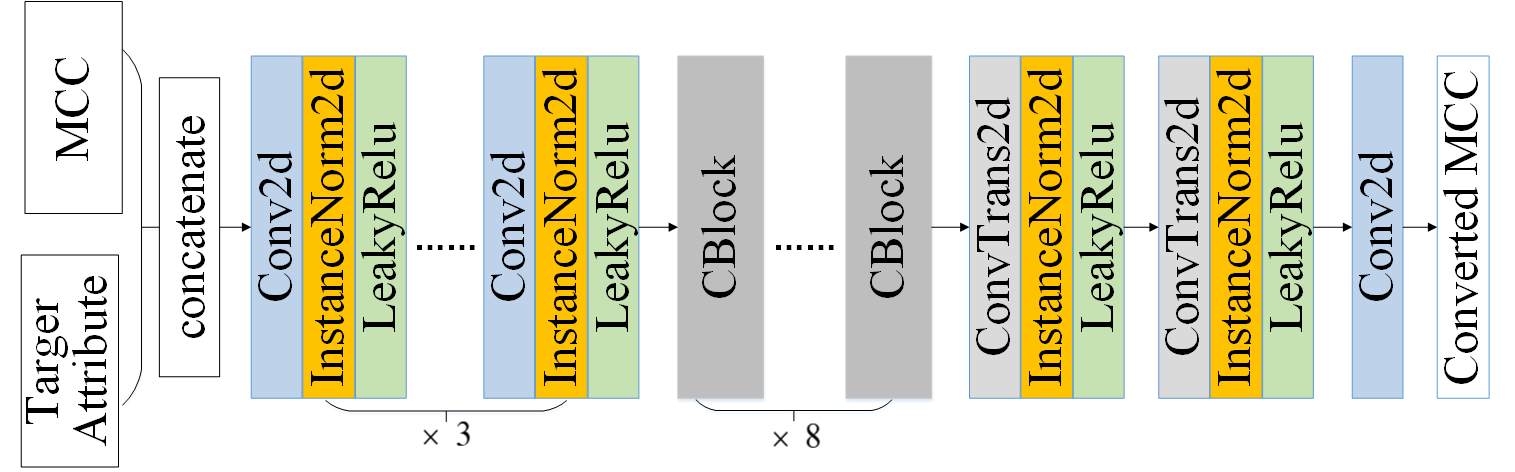}\\
  \caption{Diagram of generator network structure}
\label{fig:Generator}
\end{figure}

The network structure of the generator is shown in Figure~\ref{fig:Generator}. It is based on convolution residual network, which allows the generator to convert different length of speech. When the target attribute represented by one-hot code is expanded to the same size as the input MCC, it becomes the input of the generator after concatenating with MCC. The output of generator is the converted MCC with target attribute.

\subsubsection{Discriminator network structure}
\begin{figure}[htbp]
  \centering
  \includegraphics[width=0.85\linewidth]{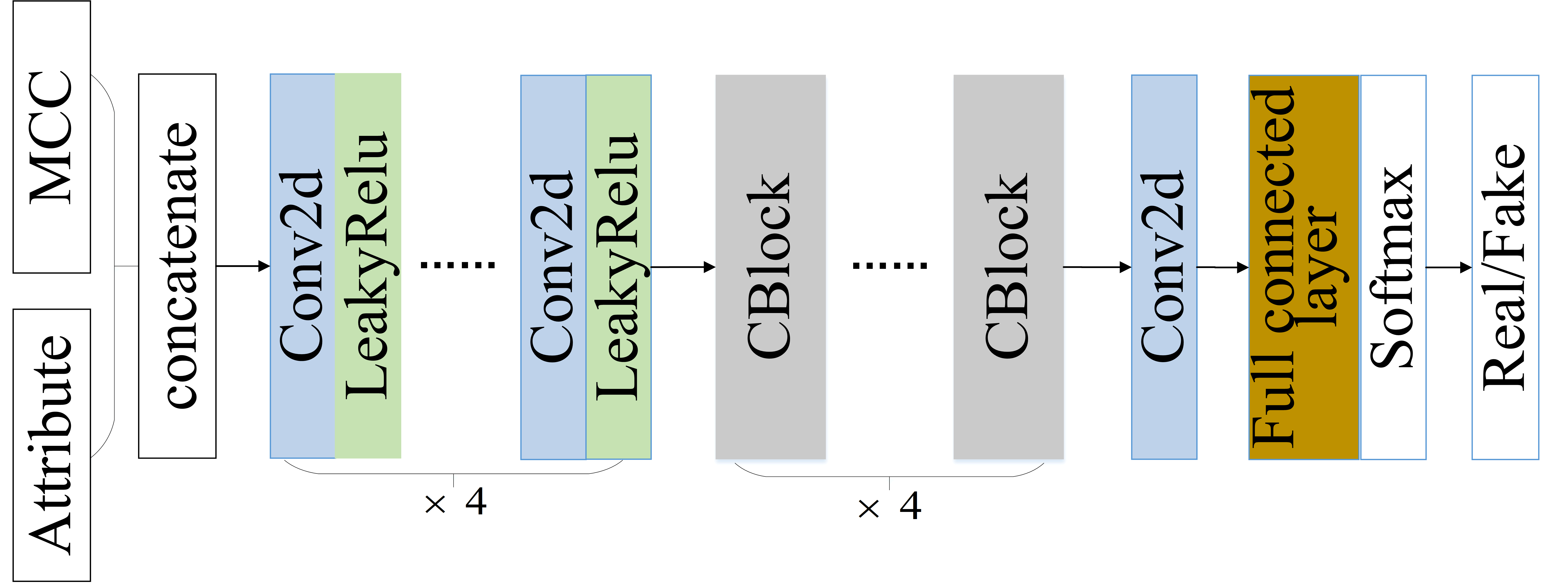}\\
  \caption{Diagram of discriminator network structure}
\label{fig:Discriminator}
\end{figure}

As show in Figure~\ref{fig:Discriminator}, we also construct the discriminator based on the convolution residual network, which allows the discriminator to discriminate different lengths of speech. The input of the discriminator is the same as the generator, and the output is the probability that weighs the likelihood that the input is the real MCC under the corresponding attribute.

\subsubsection{Classifier network structure}
\begin{figure}[htbp]
  \centering
  \includegraphics[width=0.8\linewidth]{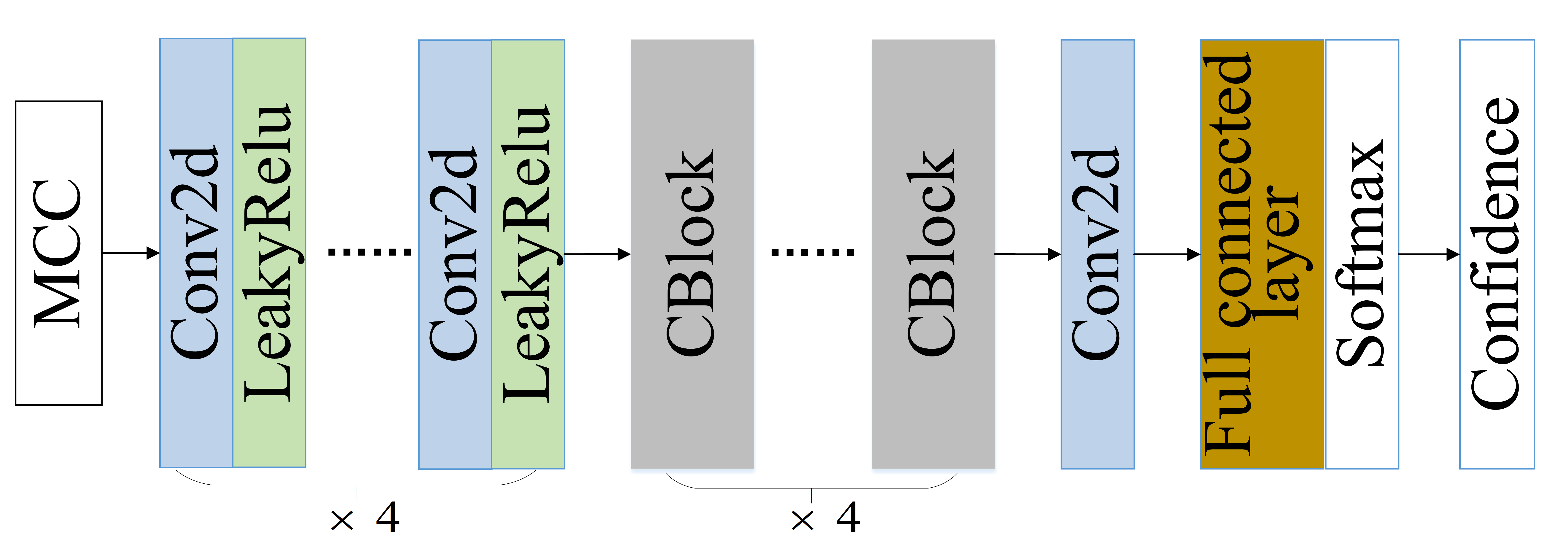}\\
  \caption{Diagram of classifier network structure}
\label{fig:Classifier}
\end{figure}

The structure of the classifier shown in Figure~\ref{fig:Classifier} is the same as the discriminator, but the input is different. The classifier takes MCC as input, and its output is the confidence of corresponding attributes.

\subsection{Speaker voice conversion module\label{Speaker VC module}}
The speaker voice conversion module makes the input speech with a specified speaker identity. Its structure is similar to the emotional voice conversion module. The difference is that the network depth of DSVC in the speaker voice conversion module is deeper than DEVC in order to avoid the feature loss. In the generator, four more CBlocks are used. And two CBlocks are adopted in discriminator and classifier respectively, which deepens the network.

In the process of multi-task speech synthesis, emotional features in speech will be lost during speaker voice conversion. According to our analysis, this is because the features extracted by speaker voice conversion model cannot fully represent the speaker's feature, that is, the feature contains a part of emotional feature. So the emotional feature is affected in speaker voice conversion process, resulting in the loss of emotional features. To address the problem, DSVC needs to extract deep speaker features. Therefore, we adopted more CBlocks to increase the depth of the model to extract deep speaker features. Because it can not only avoid the problem of gradient vanishing or explosion caused by the increase of network depth but also can extract exact deep speaker feature. The process of convolution residual network extracting features can be expressed as follow:
\begin{small}
\begin{equation}
\begin{aligned}
\begin{gathered}
F _ { l } \left( X _ { l - 1 } \right) = X _ { l - 1 } + f \left( W _ { l } , X _ { l - 1 } \right) , l \in [ 1 , N ]  \\
F _ { m } \left( X _ { m - 1 } \right) = X _ { m - 1 } , m \in [ l + 1 , N ]
\label{equation:2}
\end{gathered}
\end{aligned}
\end{equation}
\end{small}
$X _ { l - 1 }$ represents the output of the $l-1^{t h}$  layer, $W _ { l }$ represents the parameters of the $l^{t h}$ layer, and $N$ represents the number of layers of the convolution residual network. After the extraction of deep speaker feature by the convolution residual network of the $l^{th}$ layer, the layer behind $l^{th}$ layer will perform the identity mapping on the feature map. As shown in Formula~\ref{equation:2}, in fact, the network parameters $W$ behind $l^{th}$ layer is 0. This explains why the convolution residual network can extract exact deep speaker feature and ensure that the gradient will not vanish or explode.

\section{Experiments and Analysis\label{Experiments and Analysis}}
We conducted quantitative and qualitative experiments to evaluate the performance of the proposed voice conversion model and MASS framework. For quantitative experiments, we use Mel-cepstral distance (MCD)~\cite{kubichek1993mel} to measure the voice conversion effect of the proposed DEVC and DSVC. In qualitative evaluation, we conducted ablation experiments to measure the actual speech conversion effect of DEVC, DSVC and the performance of the whole MASS framework.

\subsection{Experimental platform}
The experimental platform were conducted on a server equipped with intel XEON 6420 2.6GHz $\mathbf { X }$ 18C (CPU), Tesla V100 32GiG (GPU), 16GiB DDR4-RECC 2666 (Memory), Ubuntu 16.04 (OS), Python 3.6, Pytorch-0.4.1, Keras-2.2.4.

\subsection{Datasets}
The MASS framework proposed in this paper consists of a TTS module, an emotional voice conversion module and a speaker voice conversion module, so three datasets are used to train three modules respectively. For the TTS module, we use the LJSpeech dataset~\footnote{The LJSpeech dataset can be downloaded at:~\emph{https:// keithito.com/LJ-Speech-Dataset/}} for training. The LJSpeech dataset is a public domain speech dataset consisting of 13100 short audio clips of a single speaker reading passages from 7 non-fiction books. A transcription is provided for each clip. Clips vary in length from 1 to 10 seconds and have a total length of approximately 24 hours. This dataset is widely used in TTS tasks. For speech emotional voice conversion model, EmoV-DB dataset~\footnote{The EmoV-DB dataset can be downloaded at:~\emph{https://github.com/numediart/EmoV-DB}} is used for training. The dataset contains the recording of emotional speech of four speakers. Each speech is stored in wav format with sampling rate of 16KHz. There are five emotions in the dataset: neutral, amused, sleepy, disgusted and angry. For the speaker voice conversion model, the VCTK Corpus~\footnote{The VCTK Corpus can be downloaded at:~\emph{http://homepages.inf.ed.ac.uk/jyamagis/page3/page58/page58.html}} is used for training, which includes speech data with various accents spoken by 109 native English speakers. Each speaker reads about 400 sentences, most of which are selected from newspapers and are stored in wav format at a sampling rate of 48KHz.

\subsection{Quantitative  Evaluation}
For the quantitative experiment, due to the speech synthesized by MASS contains both emotion and speaker identities, the MCD cannot assess the effect of emotional voice conversion and speaker voice conversion alone. Therefore, in quantitative experiments, the speeches synthesized by TTS module are converted into different emotional speeches and different speakers' speeches respectively, and then calculate the MCD to assess the speech synthesis effect. MCD is defined as follows~\cite{kubichek1993mel}:
\begin{small}
\begin{equation}
\begin{aligned}
\begin{gathered}
M C D = ( 10 / \ln 10 ) \sqrt { 2 \sum _ { i = 1 } ^ { 24 } \left( m c c _ { i } ^ { t } - m c c _ { i } ^ { c } \right) ^ { 2 } }
\label{equation:3}
\end{gathered}
\end{aligned}
\end{equation}
\end{small}
$ m c c _ { i } ^ { t }$ and $m c c _ { i } ^ { c }$ represent the target MCC and the converted MCC, respectively. $i$ represents the dimension of the MCC. The smaller MCD, the gap between $ m c c _ { i } ^ { t }$ and $m c c _ { i } ^ { c }$ is smaller, which represents a better voice conversion effect.

\subsubsection{Single-task emotional voice conversion experiments}
\label{Single-task emotional voice conversion experiments}

\begin{table}[htbp]
\centering
\caption{Comparison of MCD between DEVC and different emotional voice conversion methods. ``Zero effort" represents the MCD value between the original speeches and the target speeches without any voice conversion.}
\label{single-task emotional VC}
\resizebox{\linewidth}{!}{
\begin{tabular}{lccccccccccccc}
\toprule
\hline
\multirow{1}{*}
{Model} & {Natural-to-Angry} & {Natural-to-Amused} & {Natural-to-Disgusted}&{Natural-to-Sleepy}
\\
\multirow{1}{*}{DEVC}       &{\textbf{9.46}}  &{\textbf{10.26}}&  {\textbf{9.99}} & 14.40\\
\multirow{1}{*}{CycleGAN}   &11.15  &11.24&  10.08&  {\textbf{14.08}}\\
\multirow{1}{*}{StarGAN}    &14.26  &13.96&  13.91&   15.84 \\
\multirow{1}{*}{Zero effort}    &11.81  &12.90&  10.81&   11.81
\\
\hline
\bottomrule
\end{tabular}
}
\end{table}

\begin{table}[htbp]
\centering
\caption{Comparison of MCD between DSVC and different speaker voice conversion methods. ``Zero effort" represents the MCD value between the original speeches and the target speeches without any voice conversion.}
\label{single-task speaker VC}
\resizebox{\linewidth}{!}{
\begin{tiny}
\begin{tabular}{lccccccccccccc}
\toprule
\hline
\multirow{1}{*}
{Model} & {P1-to-P2} & {P1-to-P3} & {P1-to-P4}
\\
\multirow{1}{*}{DSVC}       &{\textbf{10.48}}  &{\textbf{10.34}}&  {\textbf{10.47}} \\
\multirow{1}{*}{StarGAN-VC2}   &11.84  &11.55&  11.57  \\
\multirow{1}{*}{One-shot}     &13.03  &12.67&  13.74 \\
\multirow{1}{*}{Zero effort}     &12.53  &11.62&  10.67
\\
\toprule
\multirow{1}{*}
{Model} & {P2-to-P1} & {P2-to-P3} & {P2-to-P4}
\\
\multirow{1}{*}{DSVC}       &{\textbf{10.95}}  &{\textbf{11.11}}& 10.88 \\
\multirow{1}{*}{StarGAN-VC2}   &11.74  &12.03&  {\textbf{10.80}}  \\
\multirow{1}{*}{One-shot}     &11.51  &12.05&  11.48 \\
\multirow{1}{*}{Zero effort}     &11.34  &11.62&  11.06
\\
\toprule
\multirow{1}{*}
{Model} & {P3-to-P1} & {P3-to-P2} & {P3-to-P4}
\\
\multirow{1}{*}{DSVC}       &{\textbf{10.41}}  &{\textbf{10.18}}&  {\textbf{10.62}} \\
\multirow{1}{*}{StarGAN-VC2}   &11.27  &11.14&  10.97  \\
\multirow{1}{*}{One-shot}     &12.65  &12.30&  11.77 \\
\multirow{1}{*}{Zero effort}     &10.94  &11.34&  11.35
\\
\toprule
\multirow{1}{*}
{Model} & {P4-to-P1} & {P4-to-P2} & {P4-to-P3}
\\
\multirow{1}{*}{DSVC}       &{\textbf{10.60}}  &{\textbf{10.79}}&  10.96 \\
\multirow{1}{*}{StarGAN-VC2}   &12.67  &11.81&  {\textbf{10.94}}  \\
\multirow{1}{*}{One-shot}     &12.89  &12.40&  12.23 \\
\multirow{1}{*}{Zero effort}     &10.70  &10.92&  11.59
\\
\hline
\bottomrule
\end{tabular}
\end{tiny}
}
\end{table}

As for emotional voice conversion experiments, we compare the emotional voice conversion method based on CycleGAN~\cite{zhou2020transforming} and the emotional voice conversion method based on StarGAN~\cite{rizos2020stargan} with the DEVC. Four kinds of emotion combination experiments were carried out: natural-to-angry, natural-to-amused, natural-to-disgusted and natural-to-sleepy. Firstly, 25 speeches of natural class are randomly selected from the test set of EmoV-DB dataset, and then these 25 speeches are converted into angry class, amused class, disgusted class and sleepy class by DEVC, CycleGAN and StarGAN respectively. The converted speeches are compared with the target speeches to calculate average MCD. The experimental results are shown in Table~\ref{single-task emotional VC}. Obviously, it can be found that whether the natural class speech is converted into the angry class, the amused class or the disgusted class, DEVC's MCD is smaller than the other two baselines, which indicates that the MCC of speech converted by DEVC is closer to the MCC of the target speech. This means that DEVC can generate speech with specified emotions more effectively.

However, we found that for the voice conversion natural-to-sleepy, the MCD of DEVC and its baselines are larger than those of the other three combinations. We infer that some of the audio in the sleepy class express sleepy's emotion through the sound of yawning. On the one hand, because the model manipulate MCC and F0, which cannot fit the yawning sound. On the other hand, in the sleepy class, the speech speed is much slower than the speech in natural class, and the duration is much longer than natural class. So there is a large gap between the two classes, which has a great impact on the voice conversion effect of the model. In other words, due to the dataset, neither DEVC nor its baselines can achieve a good voice conversion of natural-to-sleepy, so we will not discuss this natural-to-sleepy voice conversion experiment in the following experiments.

\subsubsection{Single-task speaker voice conversion experiments}
\label{Single-task speaker voice conversion experiments}
As for the speaker voice conversion experiments, the process is the same as the emotional voice conversion experiments, except that we randomly select test data from the test set of the VCTK corpus. We conducted all combined voice conversion experiments: P1-to-P2, P1-to-P3, P1-to-P4, P2-to-P1, P2-to-P3, P2-to-P4, P3-to-P1, P3-to-P2, P3-to-P4, P4-to-P1, P4-to-P2 and P4-to-P3. P1, P2, P3 and P4 represent 4 different speaker identities. This experiment compares the StarGAN-VC2~\cite{kaneko2019stargan} based speaker voice conversion method and One-shot~\cite{chou2019one} based speaker voice conversion method with DSVC. The single-task speaker voice conversion experiment results are shown in Table~\ref{single-task speaker VC}. It can be found that although the speaker voice conversion effect of DEVC is not as good as StarGAN-VC2 when performing P2-to-P4 and P4-to-P3 conversion, the gap with Stargan-VC2 is very small. And in other combined voice conversion experiments, DEVC can get a better voice conversion effect, which shows that the overall performance of DEVC is better than baselines.


\subsubsection{Dual-task speech synthesis experiment with specified text and emotion}
\label{Dual-task speech synthesis experiment with specified text and emotion}
The dual-task speech synthesis is to synthesize specified text and emotional speech or specified text and speaker identity speech. The difference between this experiment and the single-task voice conversion experiment is that in this experiment, the TTS module first synthesizes the speech with specified text, and the voice conversion module converts it to achieve dual-task speech synthesis, that means voice conversion module needs to convert the synthesized speech of TTS module. This is more difficult than single-task voice conversion, because there is a certain gap between the naturalness of the speech synthesized by the TTS module and the speech in the dataset. In other words the naturalness of the synthesized speech is  challenging, and the results of this part will be shown in Section~\ref{Dual-task qualitative evaluation experiments}.

As for dual-task speech synthesis experiments for specified text and emotion, firstly, 25 speeches of natural class are randomly selected from the test dataset of Emov-DB dataset. The corresponding text is obtained according to selected speeches, and then the corresponding speeches are synthesized by TTS module. Subsequently, emotion in these speeches are changed to angry, disgusted and amused. Finally, compared synthetic speeches with target speeches, the average value of MCD is calculated. In order to compare the effects of different TTS models on dual-task speech synthesis, we also compare the voice conversion effects of DEVC and baselines on TTS-tacotron2~\cite{shen2018natural} and TTS-DC~\cite{tachibana2018efficiently} which are two state-of-art TTS models. The experimental results of dual-task speech synthesis experiments are show in Table~\ref{dual-task emotional speech synthesis}. From the experimental results we can find that for speeches synthesized by the different TTS model, DEVC can always obtain a smaller MCD than the baselines, which means that DEVC can better convert the speech synthesized by the TTS model and is not affected by the TTS model. In addition, the MCD of the speech synthesized by TTS-transformer after voice conversion is smaller than TTS-tacotron2 and TTS-DC, which shows that the speech synthesized by TTS-transformer is closer to the target speech. This is the reason why TTS-transformer is selected as the TTS module in MASS framework.
\begin{table}[htbp]
\centering
\caption{Comparison of MCD value with different dual-task speech synthesis methods with specified text and emotion. ``*+Zero-effort" represents the MCD value between the speeches synthesized by TTS module and the target speeches without any voice conversion.}
\label{dual-task emotional speech synthesis}
\resizebox{\linewidth}{!}{
\begin{tabular}{lccccccccccccc}
\toprule
\hline
\multirow{1}{*}
{Model} & {Natural-to-Angry} & {Natural-to-Amused} & {Natural-to-Disgusted}
\\
\multirow{1}{*}{TTS-transformer+DEVC}       &{\textbf{10.76}}  &{\textbf{11.02}}&  {\textbf{11.21}} \\
\multirow{1}{*}{TTS-transformer+CycleGAN}   &12.00  &13.84&  12.48  \\
\multirow{1}{*}{TTS-transformer+StarGAN}     &15.84  &15.73&  15.24 \\
\multirow{1}{*}{TTS-transformer+Zero effort}     &15.98  &15.94&  15.00
\\
\toprule
\multirow{1}{*}{TTS-tacotron2+DEVC}       &{\textbf{11.58}}  &{\textbf{11.51}}&  {\textbf{11.46}} \\
\multirow{1}{*}{TTS-tacotron2+CycleGAN}   &12.51  &13.21&  12.75  \\
\multirow{1}{*}{TTS-tacotron2+StarGAN}     &14.83  &15.98&  14.76 \\
\multirow{1}{*}{TTS-tacotron2+Zero effort}     &16.49  &16.36&  15.05
\\
\toprule
\multirow{1}{*}{TTS-DC+DEVC}       &{\textbf{11.49}}  &{\textbf{11.09}}&  {\textbf{11.59}} \\
\multirow{1}{*}{TTS-DC+CycleGAN}   &13.72  &12.99&  13.76  \\
\multirow{1}{*}{TTS-DC+StarGAN}     &16.09  &15.62&  15.42 \\
\multirow{1}{*}{TTS-DC+Zero effort}     &16.75  &17.27&  16.24
\\
\hline
\bottomrule
\end{tabular}
}
\end{table}

\begin{table}[htbp]
\centering
\caption{Comparison of MCD value with different dual-task speech synthesis methods with the specified text and speaker identity. ``*+Zero-effort" represents the MCD value between the speeches synthesized by TTS module and the target speeches without any voice conversion.}
\label{dual-task speaker speech synthesis}
\resizebox{\linewidth}{!}{
\begin{tabular}{lccccccccccccc}
\toprule
\hline
\multirow{1}{*}
{Model} & {P1-to-P2} & {P1-to-P3} & {P1-to-P4}
\\
\multirow{1}{*}{TTS-transformer+DSVC}       &{\textbf{11.05}}  &{\textbf{10.59}}&  {\textbf{10.94}} \\
\multirow{1}{*}{TTS-transformer+StarGAN-VC2}   &13.03  &13.14&  12.17  \\
\multirow{1}{*}{TTS-transformer+One-shot}     &13.64  &12.80&  13.12 \\
\multirow{1}{*}{TTS-transformer+Zero effort}     &14.94  &14.30&  14.27
\\
\toprule
\multirow{1}{*}{TTS-tacotron2+DSVC}       &{\textbf{11.49}}  &{\textbf{11.61}}&  {\textbf{11.04}} \\
\multirow{1}{*}{TTS-tacotron2+StarGAN-VC2}   &12.46  &12.58&  12.93  \\
\multirow{1}{*}{TTS-tacotron2+One-shot}     &13.02  &12.38&  12.70 \\
\multirow{1}{*}{TTS-tacotron2+Zero effort}     &15.04  &15.16&  14.74
\\
\toprule
\multirow{1}{*}{TTS-DC+DSVC}       &{\textbf{11.59}}  &{\textbf{11.97}}&  {\textbf{11.05}} \\
\multirow{1}{*}{TTS-DC+StarGAN-VC2}   &12.78  &12.45&  12.57  \\
\multirow{1}{*}{TTS-DC+One-shot}     &12.95  &12.31&  12.07 \\
\multirow{1}{*}{TTS-DC+Zero effort}     &16.06  &16.25&  15.78
\\
\hline
\bottomrule
\end{tabular}
}
\end{table}

\subsubsection{Dual-task speech synthesis experiments with specified text and speaker identity}
\label{Dual-task speech synthesis experiments with specified text and speaker identity}
The experimental setup of dual-task speech synthesis with specified text and speaker identity is similar to the dual-task speech synthesis with specified text and emotion. The only difference is the dataset. Firstly, We randomly select 25 speeches from the test set of the VCTK corpus and obtain their corresponding texts. After that, according to the obtained texts, use the TTS model to synthesize speeches, and use P1 to represents the speaker identity of the speech synthesized by TTS. Finally, the speaker identities of synthetic speeches are converted to P2, P3 and P4 through speaker voice conversion model. Table~\ref{dual-task speaker speech synthesis} shows the experimental results. We can find that the results are similar to the experimental results of dual-task speech synthesis with specified text and emotion. In other words, DSVC can perform speaker voice conversion better than baselines, and can make the speech synthesized by TTS carrying the specified speaker identity.


\subsection{Qualitative experiment}
In qualitative experiments, we conducted mean opinion score (MOS) experiments and ABX experiments to evaluate the naturalness of the synthetic speech and the similarity between the synthetic speech and the target speech respectively. In order to verify the actual voice conversion effect of DEVC and DSVC, single-task and dual-task qualitative evaluation experiments are conducted. For single-task qualitative experiments, the speech in the dataset is converted through voice conversion model and qualitative evaluation experiments are performed on converted speech; for dual-tasks qualitative experiments, the speech synthesized by the TTS module is converted by voice conversion model and subjected to qualitative evaluation experiments. Finally, qualitative evaluation experiments are performed on the speech synthesized by the MASS to measure the actual speech synthesis effect. 25 subjects with good listen ability participated in all experiments.

\subsubsection{Single-task qualitative evaluation experiments}
\label{Single-task qualitative evaluation experiment results}

\begin{figure}[htbp]
  \centering
  \subfigure[Qualitative experimental results of naturalness in single-task emotional voice conversion with 95\% confidence interval. The \emph{p}-values are 0.02, 0.04 and 0.001 respectively.]{
  \includegraphics[width=0.482\textwidth]{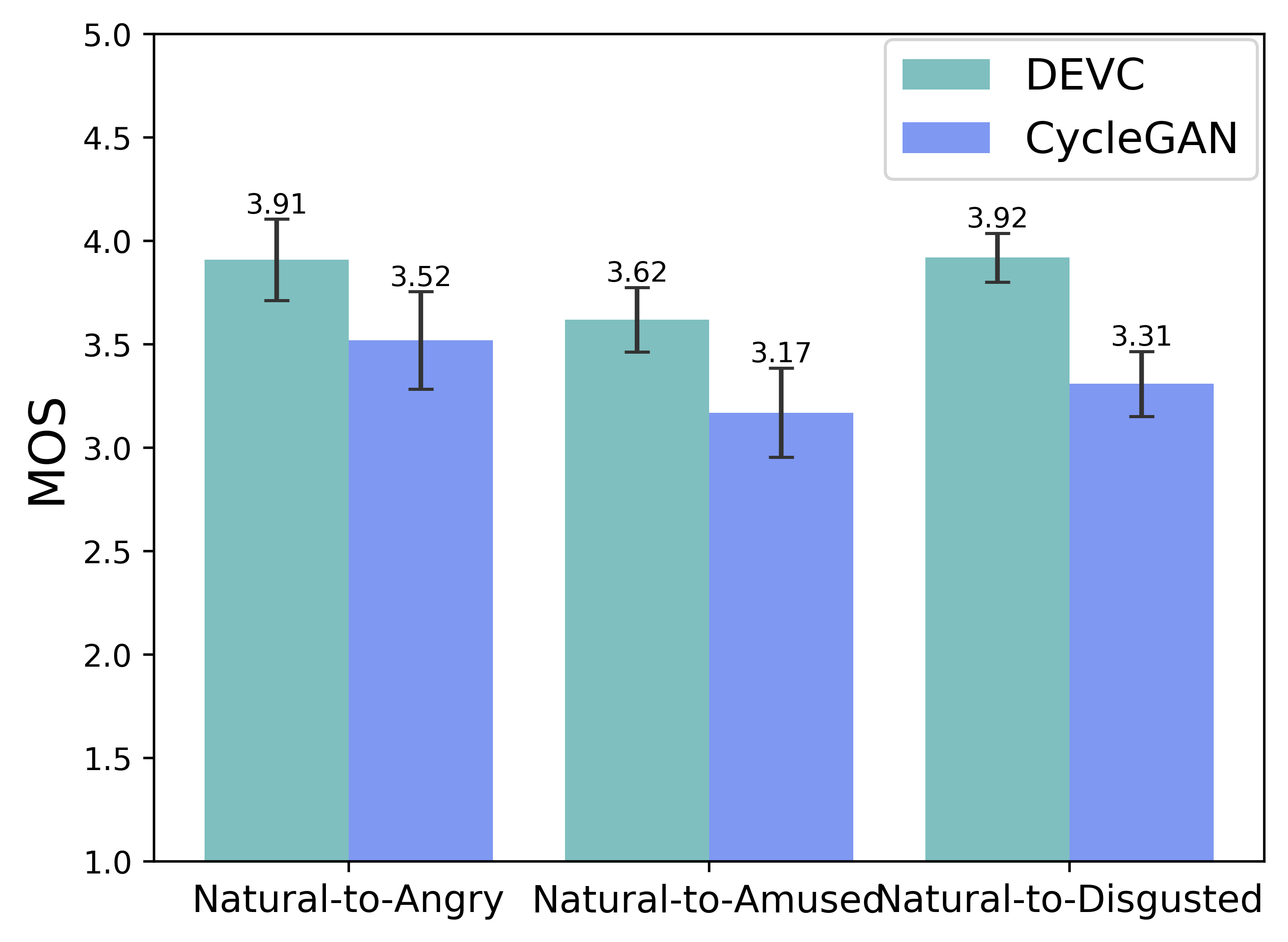}
    }
  \subfigure[Qualitative experimental results of naturalness in single-task speaker voice conversion with 95\% confidence interval. The \emph{p}-values are $9.83 \times 10 ^ { - 7 }$, $4.93 \times 10 ^ { - 9 }$ and $3.97 \times 10 ^ { - 11 }$ respectively]
  {
  \includegraphics[width=0.482\textwidth]{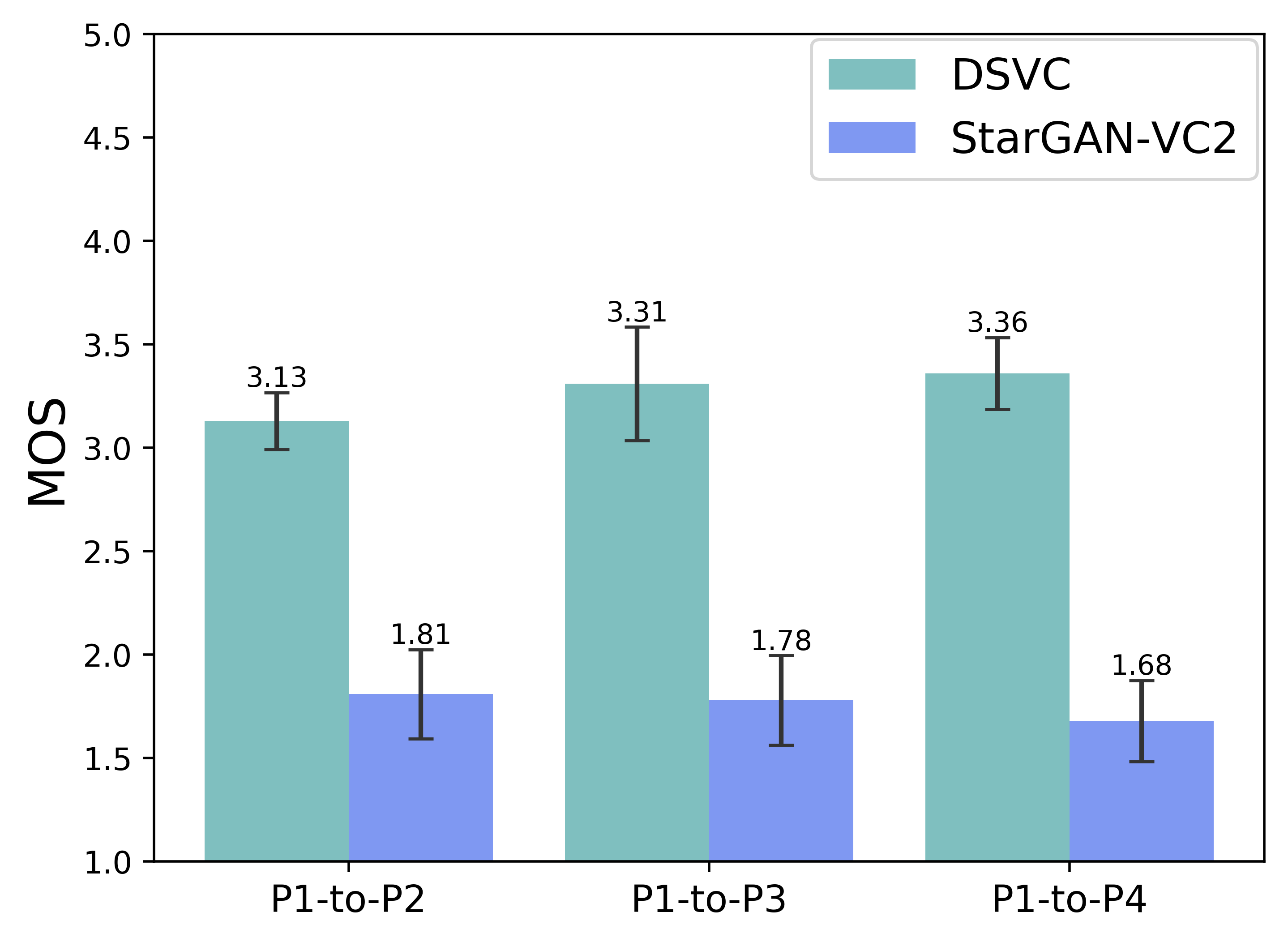}
    }
  \caption{Qualitative experimental results of naturalness in single-task voice conversion}
  \label{Subject experimental results of single-task VC}
\end{figure}
First, MOS experiment is conducted to evaluate the naturalness of the speech. In this experiment, the audience will listen to the speeches that need to be assessed, and then rate them. The score ranges from 1 to 5, 5 for the best and 1 for the worst. For score 5, the clear speech in the dataset is selected as the reference speech; for score 1, the speech that cannot be heard clearly is used as the reference speech. Each listener will listen to the reference speeches first before the test and listeners can listen to speeches any number of times until they can rate them.

We conducted MOS experiments on the speeches converted by DEVC and DSVC respectively. The baseline of DEVC is CycleGAN based emotional voice conversion method~\cite{zhou2020transforming} and DSVC's is StarGAN-VC2~\cite{kaneko2019stargan}. We regenerate the experimental data in these experiments, the generation process is the same as the single-task emotional voice conversion in Section~\ref{Single-task emotional voice conversion experiments} and single-task speaker voice conversion in Section~\ref{Single-task speaker voice conversion experiments}. The experiments of MOS are show in Figure~\ref{Subject experimental results of single-task VC}. We can find that the MOS score of DEVC and DSVC is always better than corresponding baseline, which indicate that the proposed DEVC and DSVC can better guarantee the naturalness of converted speech.

\begin{figure}[htbp]
  \centering
  \subfigure[Qualitative experimental results of similarity in single-task emotional voice conversion based on ABX experiment]{
  \includegraphics[width=0.482\textwidth]{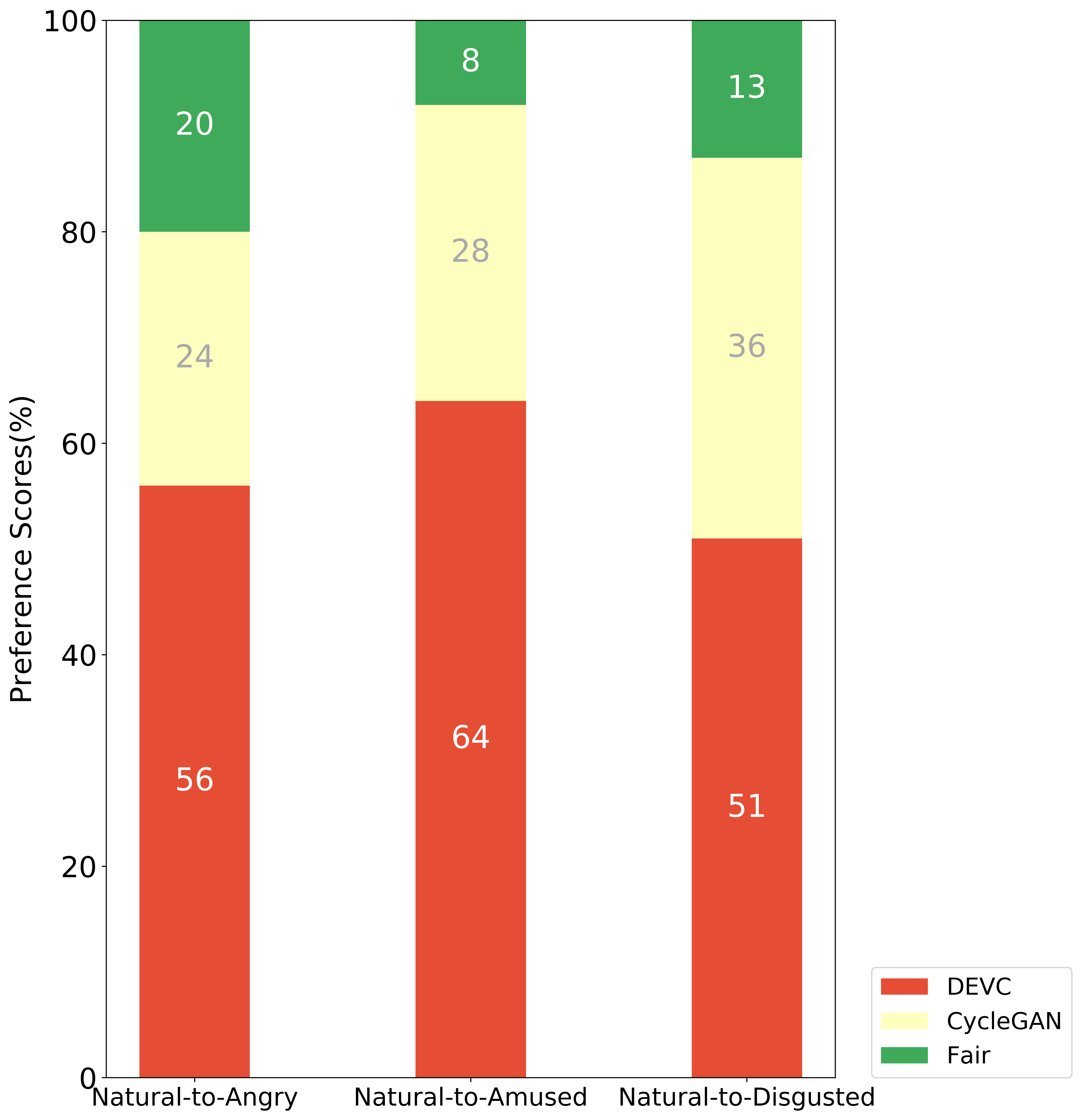}
    }
    \subfigure[Qualitative experimental results of similarity in single-task speaker voice conversion based on ABX experiment]{
  \includegraphics[width=0.482\textwidth]{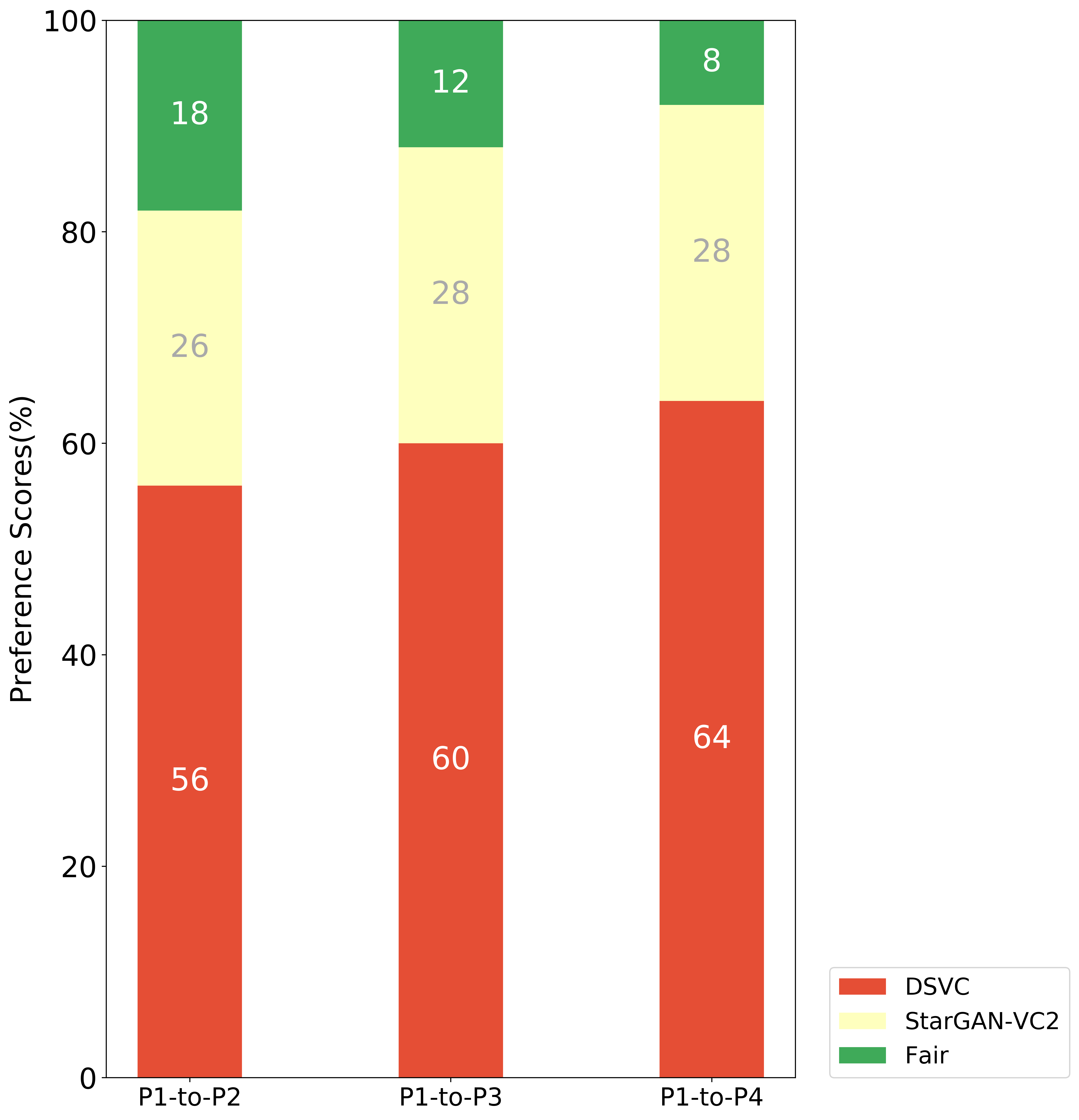}
    }
  \caption{Qualitative experimental results of similarity in single-task voice conversion based on ABX experiment}
  \label{Subject experimental results of single-task ABX}
\end{figure}
Furthermore, we conducted ABX experiments on DEVC and DSVC to evaluate the similarity between the converted speech and the target speech. As for ABX experiments, ``A" and ``B" are speeches converted by our method and baseline respectively, ``X" is the target speech. In this experiment, the listeners will listen A, B and X, and choose which of A and B is more similar to X. If they cannot tell, they will choose ``Fair". To eliminate bias, A and B are played in random order, and the listeners can listen to speeches any number of times until they can make a choice. Like the speech generation process as mentioned above MOS experiments in this section, we reproduced the speeches. Different from the above experiment, in this experiment, the listener needs to compare which speech's emotion or identity is more similar to the target emotion. The ABX experimental results of DEVC and DSVC are show in Figure~\ref{Subject experimental results of single-task ABX}. It can be found that both DEVC and DSVC can obtain higher preference scores than corresponding baseline, which indicates that DEVC and DSVC can achieve more effective voice conversion.

\subsubsection{Dual-task qualitative evaluation experiments}
\label{Dual-task qualitative evaluation experiments}

\begin{figure}[htbp]
  \centering
  \subfigure[Qualitative experimental results of naturalness in dual-task emotional voice conversion with 95\% confidence interval. The \emph{p}-values are 0.03, 0.01 and 0.004 respectively.]{
  \includegraphics[width=0.482\textwidth]{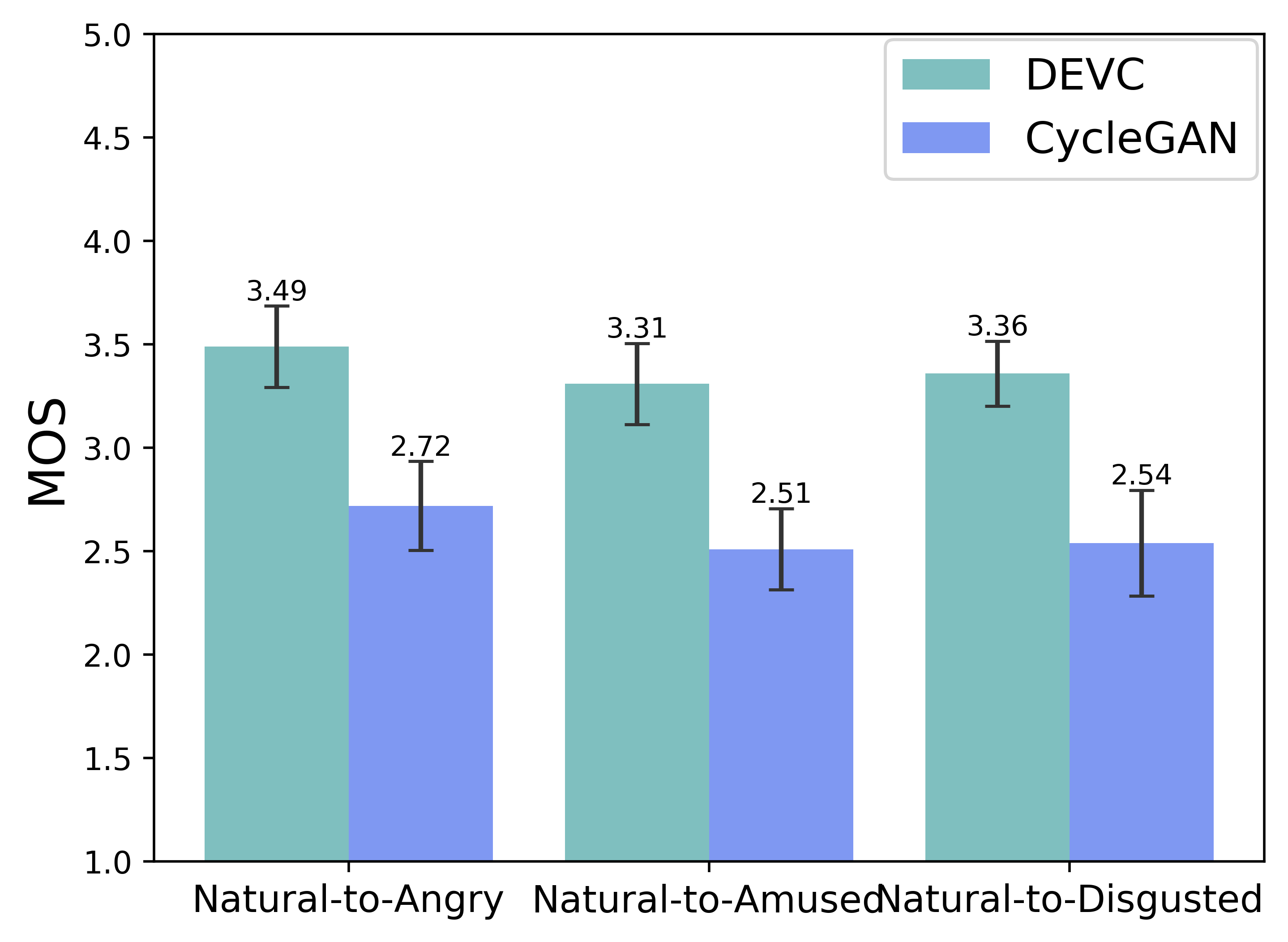}
    }
    \subfigure[Qualitative experimental results of naturalness in dual-task speaker voice conversion with 95\% confidence interval. The \emph{p}-values are $2.97 \times 10 ^ { - 7 }$, $4.90 \times 10 ^ { - 9 }$ and $3.30 \times 10 ^ { - 11 }$ respectively.]{
  \includegraphics[width=0.482\textwidth]{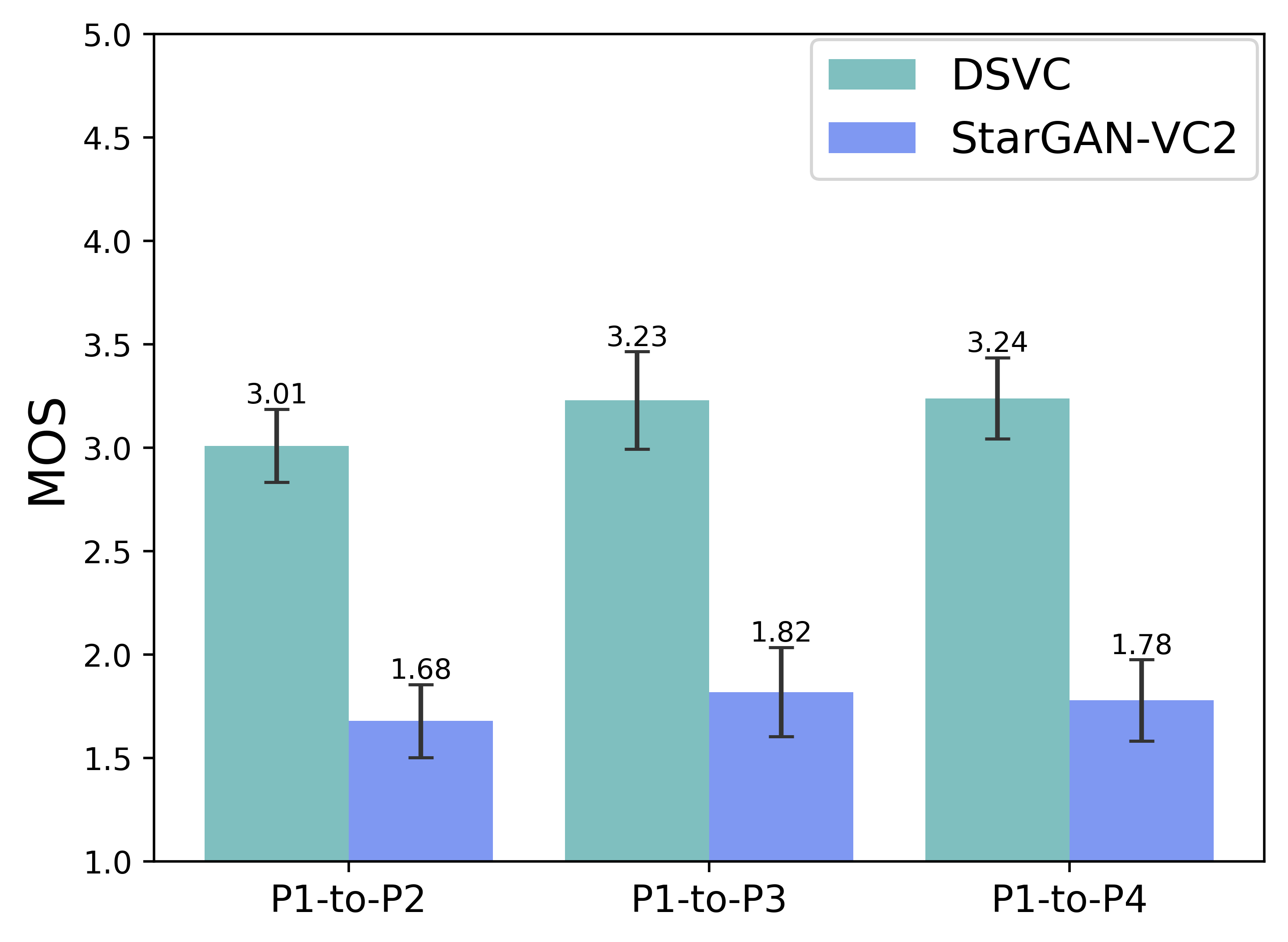}
    }
  \caption{Qualitative experimental results of naturalness in dual-task voice conversion}
  \label{Subject experimental results of dual-task MOS}
\end{figure}

For the dual-task qualitative experiment, the TTS module's speech is converted into specified emotion and speaker identity. The MOS experiment is then carried out to evaluate the naturalness of the synthesized speech and the ABX experiment to evaluate the similarity. ``A'' and ``B'' are speeches synthesized by our method and baseline; respectively, ``X'' is the target speech. We compared the dual-task qualitative evaluation experiment results with the single-task qualitative evaluation experimental results in Section~\ref{Single-task qualitative evaluation experiment results}.We re-synthesized speech for the dual-task experiment. For dual-task speech synthesis experiment with specified text and emotion, 25 speeches of natural class are randomly selected from the test dataset of the Emov-DB dataset. The corresponding text is obtained according to selected speeches, and then the corresponding speeches are synthesized by the TTS module. Subsequently, these speeches' emotions are converted into angry, disgusted, and amused emotions through different emotional voice conversion methods. In this process, we got ``A'' and ``B''. Finally, we select speeches with the exact text as the target speech X from the angry, disgusted, and amused speech dataset for the ABX experiment. The synthesis process of dual-task speech synthesis experiment with specified text and speaker is similar to dual-task speech synthesis experiment with specified text and emotion. Firstly, we randomly select 25 speeches from the test set of the VCTK corpus and obtain their corresponding texts. After that, according to the obtained texts, use the TTS model to synthesize speeches, and use P1 to represents the speaker identity of the speech synthesized by TTS. Finally, synthetic speeches' speaker identities are converted to P2, P3, and P4 through the speaker voice conversion model. Finally, we select voices with the exact text from the target speaker's test set for the ABX experiment. It should be noted that the speaker identities P2, P3, and P4 in the dual-task ABX experiment are the same as Section ~\ref{Single-task qualitative evaluation experiment results}. This is to eliminate the bias when comparing with the single-task qualitative evaluation experiment results.


The dual-task qualitative evaluation experimental results of MOS experiments are show in Figure~\ref{Subject experimental results of dual-task MOS}. Figure~\ref{Subject experimental results of dual-task MOS} shows the MOS score of DEVC and DSVC is always higher than corresponding baseline. This indicates that DEVC and DSVC convert the speech synthesized by the TTS module to be more natural and have a clearer sense of hearing. However, compared with single-task qualitative evaluation experiment results, all MOS scores have some decline. This is because there is some gap between the speech quality of TTS module and the voice quality of dataset. Nevertheless, the MOS scores of speech converted by DEVC and DSVC are not reduced much, which is basically above 3. This shows that the synthesized speech of the specified text and emotion or the speech of the specified text and speaker identity still have high naturalness.

\begin{figure}[htbp]
  \centering
  \subfigure[Qualitative experimental results of similarity in dual-task emotional voice conversion based on ABX experiment]
  {
  \includegraphics[width=0.482\textwidth]{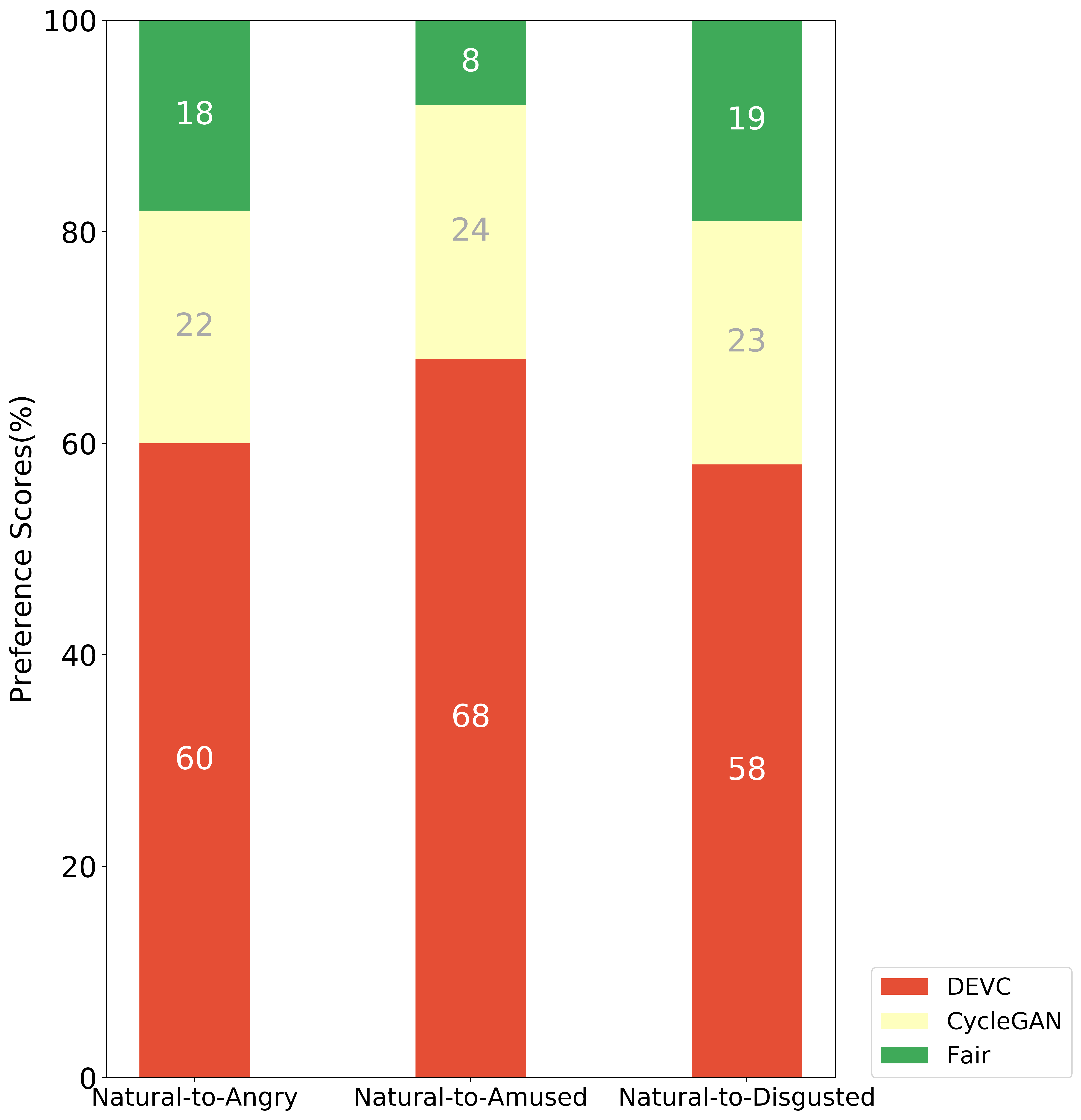}
    }
    \subfigure[Qualitative experimental results of similarity in dual-task speaker voice conversion based on ABX experiment]
    {
  \includegraphics[width=0.482\textwidth]{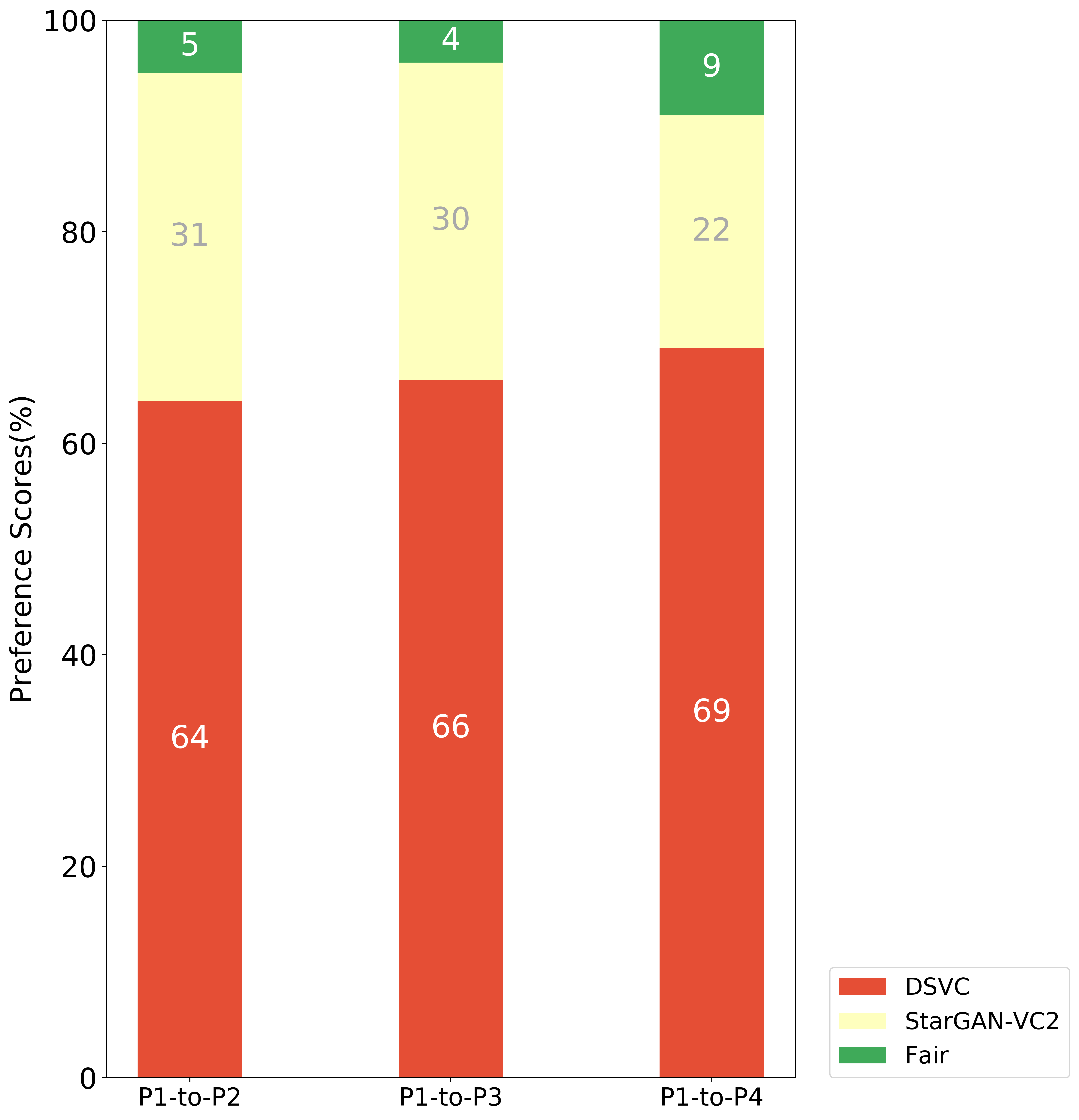}
    }
  \caption{Qualitative experimental results of similarity in dual-task voice conversion based on ABX experiment}
  \label{Subject experimental results of dual-task ABX}
\end{figure}
Figure~\ref{Subject experimental results of dual-task ABX} presents the ABX experiments. On both the similarity of emotion and the similarity of speaker identity, DEVC and DSVC can always obtain higher preference scores than the baseline which means DEVC and DSVC have a better voice conversion effect in dual-task speech synthesis. Compared with the single-task subject evaluation experiment results, it can be found that the preference scores of baselines almost decreased, which means the effect of the emotional voice conversion method based on CycleGAN and the speaker voice conversion method based on StarGAN-VC2 will be affected by the speech synthesized by TTS model. In other words, they cannot convert the speech synthesized by the TTS module effectively. However, the preference scores of DEVC and DSVC almost increase, which indicates that the voice conversion effect of DEVC and DSVC are not affected by the speech synthesized by the TTS module, and they can still effectively convert the speech synthesized by the TTS module.

\subsubsection{Qualitative evaluation of MASS}
\begin{figure}[htbp]
  \centering
  \includegraphics[width=0.5\textwidth]{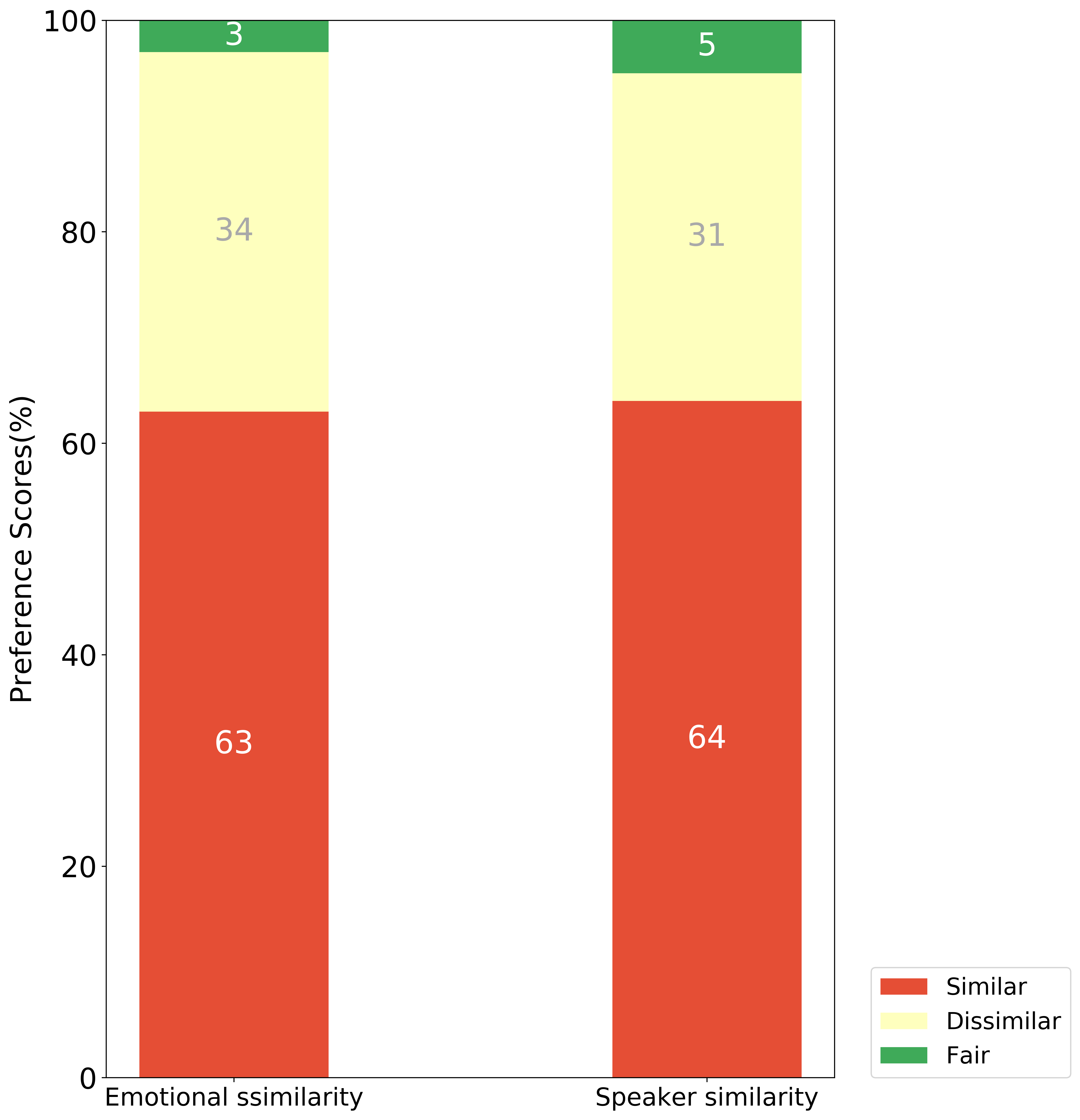}\\
  \caption{Average preference scores on emotion and speaker similarity of MASS.}
\label{MASS results}
\end{figure}

Finally, we conducted qualitative experiments on the speech synthesized by MASS. According to 15 text of natural emotional speeches that were randomly selected from the test set of the Emov-DB dataset, the TTS module synthesizes the corresponding speeches. Subsequently, emotional voice conversion makes these speeches have angry, amused and disgusted emotions respectively. Last, through speaker voice conversion, the speaker identities of P2, P3 and P4 are injected into these speeches respectively. Consequently, MASS synthesized a total of $15 * 3 * 3$ speeches. The MOS experimental result of these speeches is 3.29 which means that the sense of hearing of synthetic speeches is relatively clear. The similarity experiment between synthetic speeches and target speeches is shown in Figure~\ref{MASS results}. It can be found that more than 60$\%$ of the speeches have the specified emotions and speaker identities, which means that MASS can effectively synthesize the specified speech.

Furthermore, in order to study the influence of MASS framework on the speech synthesized by TTS module in the process of multi-task anthropomorphic speech synthesis, we explore it from  mel-spectrums and spectrums. We want to explore whether the speech synthesized by MASS framework has emotional features and identity features from mel-spectrum and spectrum respectively. The reason for analyzing emotions from mel-spectrum is that mel-spectrum is obtained on the mel-scale, and analyzing emotion in mel-spectrum is more suitable for human auditory characteristics. The human perception of frequency is non-linear. For example, when the frequency of a sound is changed from 1KHz to 2KHz, we cannot perceive the frequency of the sound to double. On the mel-scale, if the mel-frequencies of two speeches are two times different, the pitch that the human can perceive is about two times different. Therefore, if the high frequency is increased in the mel-spectrum, the emotion of the speech is likely to be transformed into an emotional emotion. As for spectrum, according to Warren's~\cite{2005Analysis} research, speech with similar spectral envelopes are considered similar. We obtain the spectrum envelope from the spectrum, and compare the spectrum envelope of the speech synthesized by MASS framework with the spectrum envelope of the target speaker's speech to explore whether the synthesized speech contains speaker's feature of the target speaker. We resynthesize some speeches randomly and draw the mel-spectrum and spectrums.

\begin{figure}[htbp]
  \centering
  \includegraphics[width=1.0\textwidth]{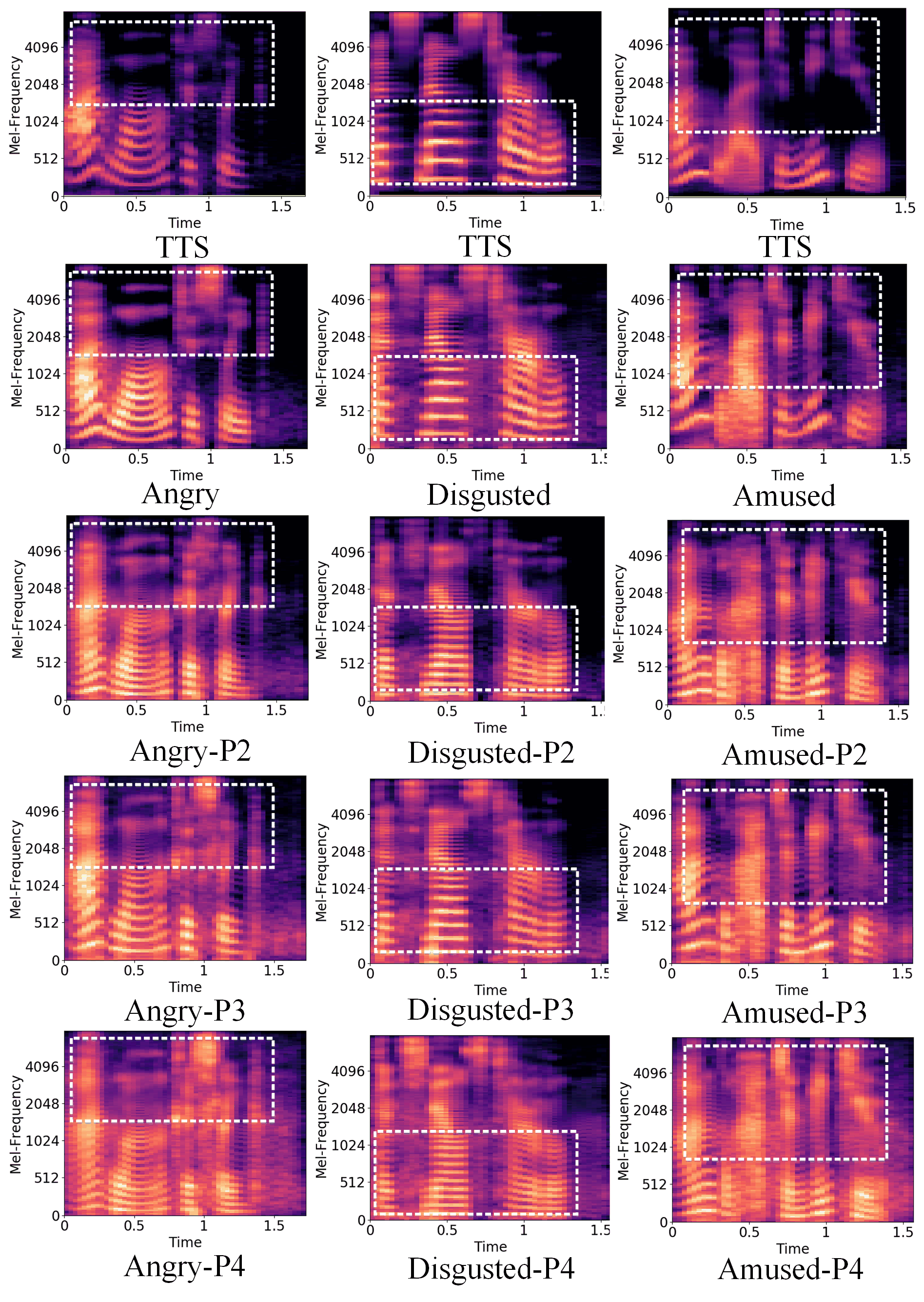}\\
  \caption{The mel-spectrum comparison of speech synthesized by MASS framework in each stage. ``TTS'' at the bottom of each figure represents the speech synthesized by TTS module, ``Angry'', ``Disgusted'' and ``Amused'' used indicate different emotions, *-* represents the speech finally synthesized by MASS framework, for example, Angry-P2 indicates that MASS synthesizes speech with angry emotion and P2 speaker identity}
\label{waveform-comparison}
\end{figure}

The mel-sprectrum of speeches synthesized by TTS module and the mel-sprectrum synthesized by MASS framework are shown in the Figure~\ref{waveform-comparison}. From the first column in Figure 12, what we can find is that when the emotion of the speech synthesized by the TTS module is converted to the angry class, the color of the high frequency part in the mel-spectrum is increased. In order to facilitate observation, we have marked this part with a white dashed box. This shows that after emotional voice conversion, the energy of the high-frequency part of the speech synthesized by the TTS module is increased, and the converted angry speech will sound louder and sharper. This corresponds to how people increase the volume and frequency of their voices to express their angry emotion. Comparing the speeches synthesized by the MASS framework with the converted angry speech and the speech synthesized by TTS, it can be found that the energy of the high frequency part of the speeches synthesized by the MASS framework is similar to that of the converted angry speech, and they are both higher than the speech synthesized by the TTS module. This shows that the emotional features of the speech after the speaker voice conversion still exist. Compare the mel-spectrum in the white dashed box in the second column of Figure~\ref{waveform-comparison}, we can find that when the emotion of the speech synthesized by TTS is converted into disgusted emotion, the energy of the low frequency part increases, and the sound sounds deeper which corresponds to people expressions of disgusted emotion in reality life. And further observation shows that the low-frequency energy of the converted disgusted speech is similar to the speeches synthesized by MASS framework, and both are higher than the speech synthesized by TTS, which shows that the speech synthesized by the MASS framework has the feature of disgusted emotion. As for the third column in Figure~\ref{waveform-comparison}, it can be found that whether it is converted amused speech or speech synthesized by MASS framework, its mel-spectrum is brighter than the speech synthesized by TTS, that is to say, the energy of converted amused speech and speech synthesized MASS framework is higher in each frequency band. Further observation shows that the energy of the high frequency part increases more than that of the low frequency part. This shows that the converted amused speech and speech synthesized by MASS framework is louder and higher pitch than the natural speech. When people express amused emotions, they usually make their voice louder and higher pitch, so the speech synthesized by MASS can express amused emotion. To summarize the above, on the one hand, this shows that DEVC can effectively synthesize speech with specified emotions. On the other hand, this also shows that DSVC will not cause the loss of emotional features after speaker voice conversion of emotional speech.

\begin{figure}[htbp]
  \centering
  \includegraphics[width=1.0\textwidth]{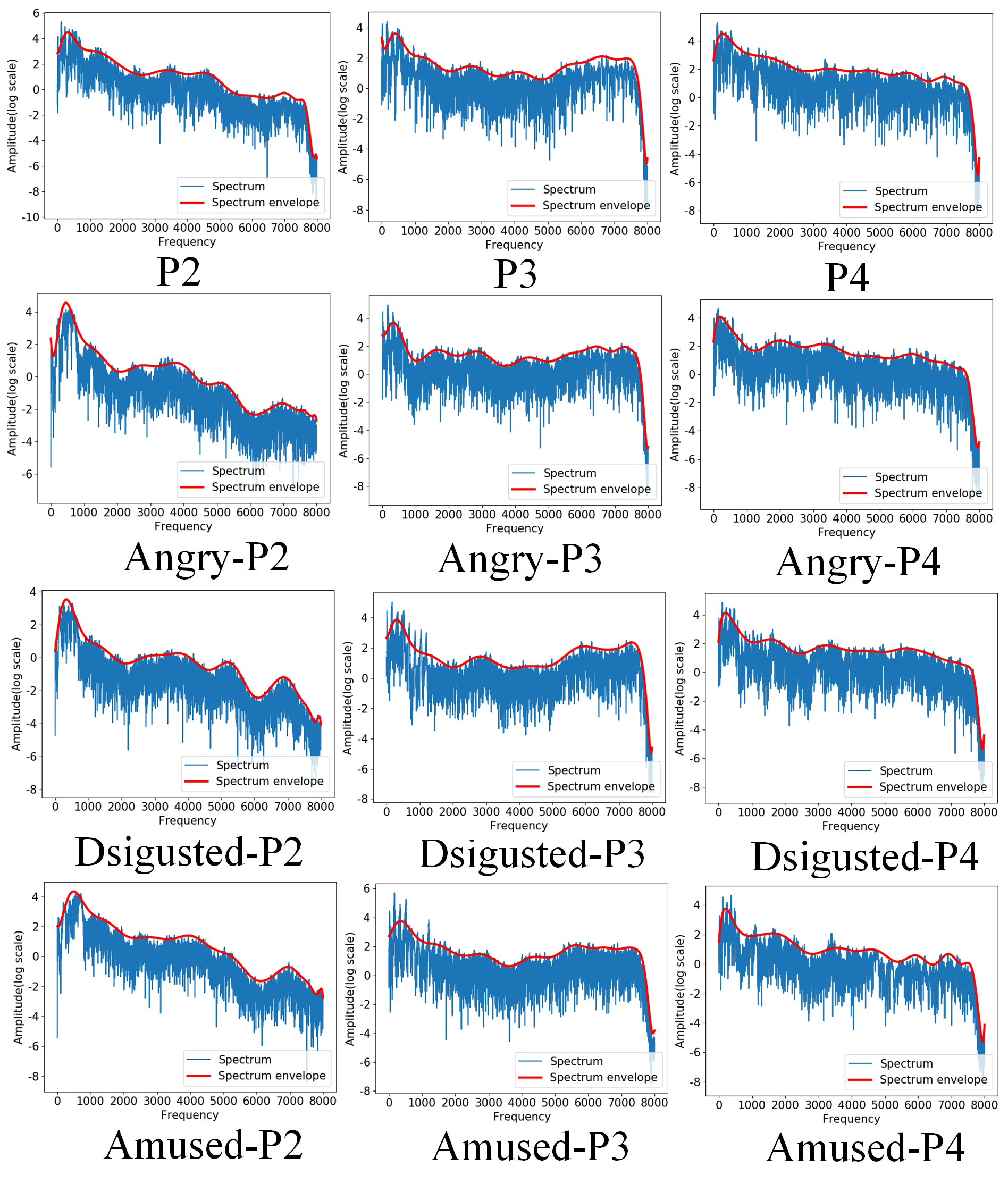}\\
  \caption{Comparison of spectrum envelope of speech synthesized by MASS and target speaker's speech. ``P2'', ``P3'' and ``P4'' at the bottom of corresponding figure represent three different target speakers. *-* represents the speech finally synthesized by MASS framework, for example, Angry-P2 indicates that MASS synthesizes speech with angry emotion and P2 speaker identity.}
\label{spectrum-comparison}
\end{figure}

However, only based on the mel-spectrum can only analyze that there is a specified emotion in the synthesized speech, it cannot analyze the speaker's identity information from the mel-spectrum. According to Warren's~\cite{2005Analysis} research, sounds with similar spectrum envelopes are considered similar. Therefore, we can compare whether the spectrum envelope of the speech synthesized by MASS framework is similar to the spectrum envelope of the target speaker's speech, and then we can analyze whether the speech synthesized by MASS framework has the specified speaker feature. Specifically, we drew the spectrums of the target speakers speeches and the spectrums of speeches synthesized by MASS framework, and obtained the spectrum envelopes according to the spectrums which are shown in Figure~\ref{spectrum-comparison}. It can be found from Figure~\ref{spectrum-comparison} that the spectrum envelopes of each column are roughly similar. Because the synthesized speech by MASS framework also has emotional features, the spectrum envelope of the speech synthesized by MASS framework will not be exactly the same as the target speaker, but the general trend of the spectrum envelope of the speech synthesized by MASS framework is similar to the target speaker and the frequency corresponding to the formant is also close. Therefore MASS framework could synthesize the speech with specified speaker indetities. Based on the above analysis of mel-spectrum and spectrum envelope, we can infer that MASS framework can effectively synthesize speech with specified emotions and specified speaker identities.

\section{Conclusions and Future Work\label{Conclusion}}
We propose MASS, a anthropomorphic speech synthesis framework that can synthesize speech with specified text, emotion and speaker identity. We conduct qualitative and quantitative experiments to evaluate the performance of the proposed DEVC, DSVC and MASS frameworks. Our experimental results show that DEVC and DSVC can effectively convert speech and the voice conversion effect will not be affected by the speech synthesized through TTS model. The qualitative evaluation experiment on MASS show that MASS can effectively synthesize target speeches.

However, the MASS framework also has some problems. Although the MASS framework integrates the TTS, DEVC and DSVC modules in series to avoid some problems, this makes the training cost of the MASS framework too high. Moreover, the DEVC and DSVC we constructed cannot transform the duration, and the effect of the generated speech can be further improved. Therefore, in future work, we will explore how to improve the training efficiency of the MASS framework and further improve the overall performance.

\section{Acknowledgement\label{Acknowledgement}}
This research was supported by the National Natural Science Foundation of China under Grant No. 62072406, the Natural Science Foundation of Zhejiang Provincial under Grant No. LY19F020025, the Major Special Funding for "Science and Technology Innovation 2025" in Ningbo under Grant No. 2018B10063.

\bibliography{bibfile}
\end{document}